\documentclass[a4paper,fleqn,usenatbib]{mnras}
\usepackage{newtxtext,newtxmath}
\usepackage[T1]{fontenc}
\usepackage{ae,aecompl}
\usepackage{graphicx}	
\usepackage{amsmath}	
\usepackage{amssymb}	

\title[Clustering Clusters]{Clustering clusters: unsupervised machine learning on globular cluster structural parameters}

\author[Pasquato et al.]{
Mario Pasquato,$^{1}$\thanks{E-mail: mario.pasquato@oapd.inaf.it}
Chul Chung,$^{2}$
\\
$^{1}$ INAF, Osservatorio Astronomico di Padova, vicolo dell'Osservatorio 5, 35122 Padova, Italy\\
$^{2}$ Center for Galaxy Evolution Research, Yonsei University, Seoul 03722, Republic of Korea\\
}

\date{Accepted XXX. Received YYY; in original form ZZZ}

\pubyear{2018}

\begin{document}
\label{firstpage}
\pagerange{\pageref{firstpage}--\pageref{lastpage}}
\maketitle

\begin{abstract}
Globular Clusters (GCs) have historically been subdivided in either two (disk/halo) or three (disk/inner-halo/outer-halo) groups based on their orbital, chemical and internal physical properties. The qualitative nature of this subdivision makes it impossible to determine whether the natural number of groups is actually two, three, or more. In this paper we use cluster analysis on the $(\log M, \log \sigma_0, \log R_e, [Fe/H], \log | Z |)$ space to show that the intrinsic number of GC groups is actually either $k=2$ or $k=3$, with the latter being favored albeit non-significantly. In the $k=2$ case, the Partitioning Around Medoids (PAM) clustering algorithm recovers a metal-poor halo GC group and a metal-rich disk GC group. With $k=3$ the three groups can be interpreted as disk/inner-halo/outer-halo families. For each group we obtain a medoid, i.e. a representative element (NGC $6352$, NGC $5986$, and NGC $5466$ for the disk, inner halo, and outer halo respectively), and a measure of how strongly each GC is associated to its group, the so-called silhouette width. Using the latter, we find a correlation with age for both disk and outer halo GCs where the stronger the association of a GC with the disk (outer halo) group, the younger (older) it is. 
\end{abstract}


\section{Introduction}
Globular clusters (GCs), being the oldest known stellar systems in the local Universe and among the best studied, are key to unraveling the formation history of our Galaxy. As a relic of the beginnings of the Milky Way, GCs are natural tools for Galactic archeology \citep[][]{2018arXiv181104798R}. One of the open questions of Galactic archeology is the accreted~/~in situ dichotomy. Another dichotomy may naturally arise in the formation of any stellar system if there are multiple bursts of star formation, resulting in populations that differ by age and metallicity. If indeed some GCs manage to undergo core collapse in their lifetime they may have clear physical differences that set them apart as a separate class with respect to pre-core-collapse GCs \citep[see e.g.][]{2013A&A...554A.129P}. Another example of three-pronged classification of GCs (dynamically young, intermediate-age, old) by their dynamical evolutionary status is introduced in the series of papers by \cite{2012Natur.492..393F, 2014ApJ...795..169A, 2015ApJ...799...44M, 2018ApJ...867..163P} based on the effects of dynamical friction on blue straggler stars.

For these and many other reasons it is natural to look for subgroups of GCs. 

In the context of an already established disk/halo dichotomy \citep[][]{1962ApJ...136..748E, 1978ApJ...225..357S}, \cite{1985ApJ...293..424Z} subdivided GCs in disk and halo families based on metallicity and spatial distribution with respect to the plane of the Galaxy. His results are based on a simple cut in metallicity (at [Fe$/$H] $= -0.8$) which separates GCs into two families: halo GCs, isotropically distributed around the Galaxy and metal-poor, and disk GCs, flattened near the Galaxy disk with a scale height of $\approx 500$pc and relatively more metal rich.
Later \cite{1988csa..proc..149L} interpreted differences in the second parameter phenomenon \citep[][]{1967AJ.....72...70V, 1967ApJ...150..469S} as an age spread among halo GCs; similarly \cite{1993ASPC...48...38Z} split the halo GC family into old halo and younger halo based on Horizontal Branch (HB) morphology and age, resulting in an overall trichotomy. This split was aimed at dividing halo GCs in two groups that followed different relations between HB morphology and metallicity, suggesting that the second parameter of HB morphology \citep[see][]{1994ApJ...423..248L, 2010ApJ...708..698D} is related to the environment of GC formation.
Like in \cite{1985ApJ...293..424Z}, the groups proposed by \cite{1993ASPC...48...38Z} are defined using cutoffs based on educated guesses supported by subject matter knowledge rather than through an automated procedure. This makes it impossible to check quantitatively whether there are three, rather than two, or four, or $N$ groups of GCs by measuring the quality of our grouping.

Only a much later paper by \cite{2009MNRAS.398.1706F} (F09 in the following) would work along the lines of \cite{1985ApJ...293..424Z}, i.e. looking for GC families that could act as a second parameter, but using a reproducible, objective method. In particular, F09 use cladistics methods adapted from biology to cluster GCs in the variable space defined by relative ages, metallicity, absolute V magnitude, and maximum effective temperature on the HB. They present their results by means of a dendrogram which bears a strong similarity to a phylogenetic tree of biological species. In the following (in particular in Sect.~\ref{comparison}) we will mainly concentrate on the three groups of GCs obtained by splitting the F09 evolutionary tree at its three main branches, even though the tree structure encodes more information about the GC clustering structure than the groups alone. One notable example of this is the fact that the group identified by F09 as outer halo splits off first at the tree root, while the remaining two groups (inner halo and disk) are divided at a later branching, suggesting that they are more similar to each other than to the former.

In addition to F09, there are few other examples of attempts at clustering GCs using automated procedures. An early example is \cite{1989SvA....33..280E}, which also obtains a dendrogram, but using agglomerative clustering methods. A subsequent work also using agglomerative clustering is \cite{1993A&A...270...83C}.
Agglomerative clustering, not unlike the cladistics approach of F09, traverses a dataset's clustering structure at all scales. It works by building a hierarchy where individual data points are merged with their nearest neighbor, subsequently merging clusters that are progressively further from each other, until all data is gathered in a single cluster. Given this hierarchy, which can be represented by a dendrogram, it is up to the researcher to split it at a chosen depth, obtaining two, three, or more clusters. This element of arbitrarity, combined with the fact that most such algorithms are based on a greedy approach (finding the best cluster merge at a given level of the hierarchy rather than the best global subdivision into groups), makes it hard to use agglomerative clustering and similar methods to find an optimal number of clusters and optimal clustering structure at a given number of clusters.
Similar criticism was leveled at hierarchical clustering approaches by \cite{2007A&A...472..131C} (C07 in the following), who used a partitioning approach instead. In this paper we follow C07 in using a partitioning algorithm, but take a radically different approach in the choice of the variable space we work with, more along the lines of F09.

\section{Methods}
We use unsupervised machine learning (clustering) to subdivide GCs into $k$ groups in a more reproducible and less arbitrary way than could otherwise be done by eye. However, with respect to previous studies we take the following steps to ensure that the results of our cluster analysis are physically meaningful:

\begin{itemize}
\item we choose a limited, manageable number of observable GC parameters: mass, central velocity dispersion, half-light radius, metallicity, and height on the Galactic plane. The number of GCs in the Galaxy is only of order $\approx200$ ($110$ in our sample), so it is vital to limit the dimensionality of the variable space to avoid incurring in the so-called \emph{curse of dimensionality};
\item we use parameters obtained from a homogeneous catalog rather than a compilation (except for [Fe$/$H]), so we do not end up finding clustering structure that is due to the piecemeal way the data was acquired rather than intrinsic to the data;
\item we consider only GC parameters that are not based on arbitrary definitions\footnote{this is admittedly an overstatement: for example the masses we use are model dependent, using the half-light radius instead of the core radius is a rather arbitrary choice, and so on. However the fact that e.g. a mass is double another has certainly a clear physical interpretation and dynamical/evolutionary consequences; this is not so clear cut with e.g. HB morphology.}: mass, half-light radius, velocity dispersion, distance from the Galactic plane, and metallicity, as opposed to $B - V$ color, HB morphology, etc. sometimes used in previous studies. The variables we chose are called \emph{ratio variables} in statistics, and are variables for which the $0$ point of the scale is not arbitrary;
\item we take the logarithm of our adopted variables (except for [Fe$/$H], which is already logarithmic) so that changes in measurement units do not affect our results, as they amount to a rigid shift of all data points which does not affect distances in parameter space; contrast this with previous studies using e.g. core radii measured in parsec, and galactocentric distance measured in kiloparsec;
\item we do not otherwise standardize our variables, as their ranges are all comparable;
\item finally, we study the effect of changing the metric used for clustering from plain Euclidean to a Mahalanobis distance that weighs some coordinates more than others.
\end{itemize}

As opposed to C07 we use Principal Component Analysis (PCA) only to visualize our results, rather than to select the parameters to include in our study. We include mass, projected half-light radius, and velocity dispersion \citep[the three \emph{fundamental plane} variables; see e.g.][]{2008A&A...489.1079P} but no additional structural parameters because GC surface brightness profiles are acceptably described by three-parameter models and further parameters may introduce collinearities, which would amount to weighing a given coordinate twice (or more, depending on the number of collinear variables) when computing distances. In fact, precisely because of the fundamental plane relation, the three parameters we chose may already be somewhat redundant. 

\subsection{Partitioning Around Medoids}
Following C07 we use a partitioning method rather than a hierarchical method. We also try to determine the optimal number of clusters objectively rather than arbitrarily. We chose to use the Partitioning Around Medoids (PAM) algorithm \citep[][]{kaufman1987clustering}.
The reasons for this choice are:
\begin{itemize}
\item PAM returns a representative item for each group it finds, known as a medoid. Properties of the groups found by PAM can, to some extent, be summarized by the properties of the respective medoid.
\item PAM can be made more robust by changing the distance it uses to calculate dissimilarity between points, a topic we reserve for discussion in future papers
\item PAM is conveniently implemented in the \emph{cluster} library in R \citep[][]{clusterlibrary}
\end{itemize}

The PAM algorithm is an alternative to the $K$-means algorithm \citep[][]{steinhaus1956division, ball1965isodata, macqueen1967some, lloyd1982least}. A set of objects (in our case GCs) is represented as points in a multidimensional space. A suitable distance (e.g. Euclidean distance) is used to measure the dissimilarity of points in this space. PAM works by iterating through a sequence of steps intended to minimize the average dissimilarity of points to their groups medoid by either picking a different medoid or switching objects to a different group. An in-depth description of PAM and other clustering algorithms, including some agglomerative methods discussed previously, can be found in the comprehensive book by \cite{KR}.

\subsection{Silhouettes}
\cite{ROUSSEEUW198753} introduced silhouettes to quantify how well a given data point fits in a clustering structure. Given clusters defined on a data set, so each data point is uniquely assigned to a cluster, the silhouette method can be used to assess the quality of the clustering irrespective of how the clusters were determined in the first place. While we will calculate silhouettes for clusters obtained using the PAM algorithm, they could in principle be used on clusters determined by any other method, including by eye.
Given $N$ data points $d_1$, ..., $d_N$ $\in S$ partitioned into $k$ clusters with $k$ such that $N > k > 1$, we will consider one such cluster $A$ containing $N_A > 1$ points, and a point $d_i \in A$. Let $D: S \times S \to \mathbb{R}_0^{+}$ be a distance on $S$, and define
\begin{equation}
a_i = \frac{1}{N_A - 1} \sum_{j \ne i, d_j \in A} D(d_i, d_j).
\end{equation}
The quantity $a_i$ is the average distance of point $d_i$ from all of the other points in the same cluster, A.
Now let us consider a cluster $C \ne A$ with $N_C$ elements and define
\begin{equation}
c_i = \frac{1}{N_C} \sum_{j, d_j \in C} D(d_i, d_j).
\end{equation}
The quantity $c_i$ represents the average distance of point $d_i$ from the points in cluster $C$. We can then consider the minimum of this average distance over all the clusters that are not A:
\begin{equation}
b_i = \min_{C \ne A} c_i
\end{equation}
With a finite number of clusters $k$ this minimum exists, but it is not guaranteed that there is only one cluster $B$ realizing it. However in practice it is very unlikely that two or more clusters have exactly the same average distance to point $d_i$, so we can assume there is only one and call it the \emph{neighbor} cluster of $d_i$. This is in a sense the cluster to which $d_i$ should have been assigned had it not been assigned to $A$. It is intuitive that $a_i \ll b_i$ means that $d_i$ is in the average much nearer to members of cluster $A$ than of cluster $B$, so it is clustered properly. On the other hand $a_i > b_i$ means we would be better off grouping $d_i$ with cluster $B$.
The \emph{silhouette width} for point $d_i$ is defined as
\begin{equation}
s_i = \frac{b_i - a_i}{\max{a(i), b(i)}}.
\end{equation}
If $a_i > b_i$ then $s_i > 0$, and vice-versa. Also $-1 < s_i < 1$.
Silhouette widths can be used for several purposes. For a given point $d_i$ we can use its $s_i$ to express how well it is clustered, i.e. is it placed firmly inside of the assigned cluster ($s_i \simeq 1$), on the fringe ($s_i \simeq 0$), or even misclassified ($s_i \simeq -1$)? For a cluster $C$ we can consider the average of the $s_i$ over the cluster to determine how cohesive the cluster is (i.e. is the cluster made up of rubble that would be better off assigned to other, neighboring clusters? Or is it mostly made of well classified points?), and for the whole dataset we can use the average $s_i$ to measure the quality of the overall clustering structure, which can be useful in comparing structures with different numbers of clusters and ultimately in picking an optimal number of clusters $k$.

\section{Data}
We obtain state of the art structural parameters (mass, sky-projected half-light radius, and central velocity dispersion) for $112$ Milky Way GCs from \cite{2018MNRAS.478.1520B}, [Fe$/$H] metallicities from \cite{1996AJ....112.1487H} (updated 2010), and distances from the Galactic plane from \cite{2019MNRAS.482.5138B}. Our combined dataset contains $110$ GCs.

\section{Results}
\begin{figure}
\includegraphics[width = 0.95\columnwidth]{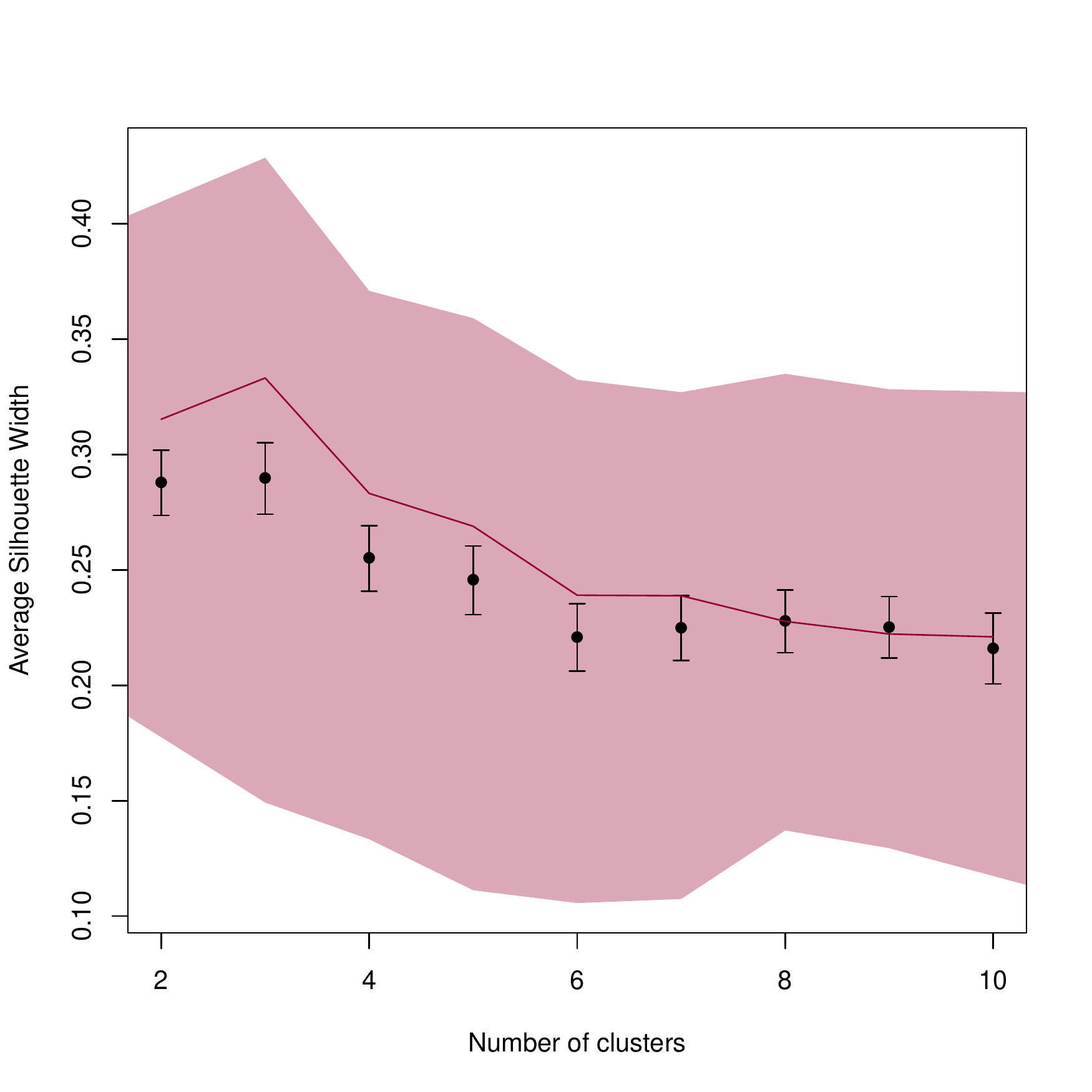}
\caption{Average silhouette width (filled points, error bars are one standard deviation of the mean) as a function of the number of clusters, $k$. The wine red shaded area represents the first and third quartiles of the silhouette width sample, and the continuous line the median. The optimal number of clusters ($k = 3$) maximizes the average and the median silhouette width, even though the difference of the former from the $k = 2$ case is non-significant, making $k = 2$ also a reasonable choice for the number of clusters. \label{Optimal}}
\end{figure}

\begin{figure}
\includegraphics[width = 0.95\columnwidth]{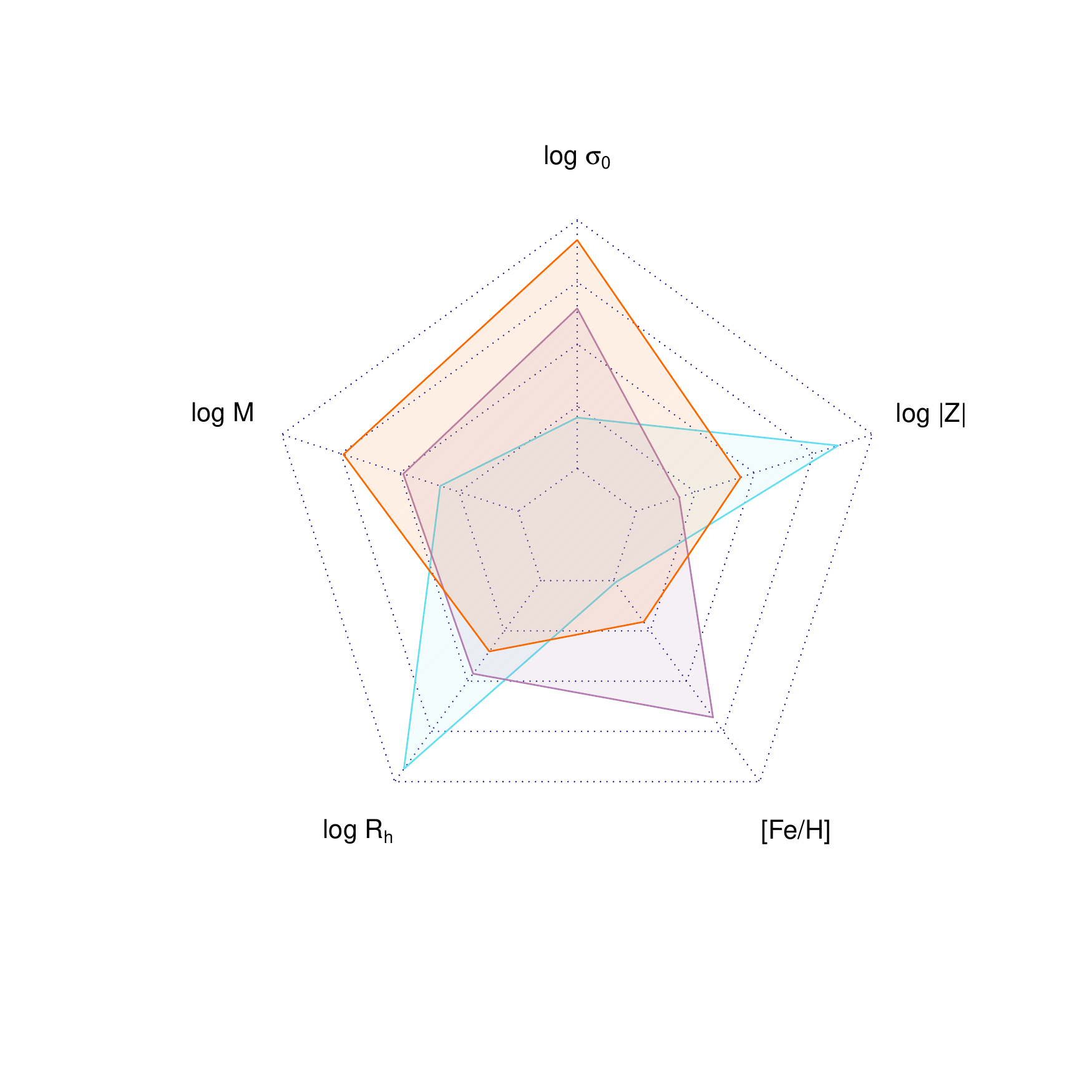}
\caption{Group medoid characteristics for the three clusters shown on a spider plot. NGC $6352$, the medoid of the disk group, is shown in violet; NGC $5986$, the medoid of the inner halo group in orange; and NGC $5466$, the medoid of the outer halo group, in light blue. All variables were centered and rescaled to span the $[0,1]$ range for the purposes of this display.\label{medoidspider}}
\end{figure}

\begin{figure}
\includegraphics[width = 0.95\columnwidth]{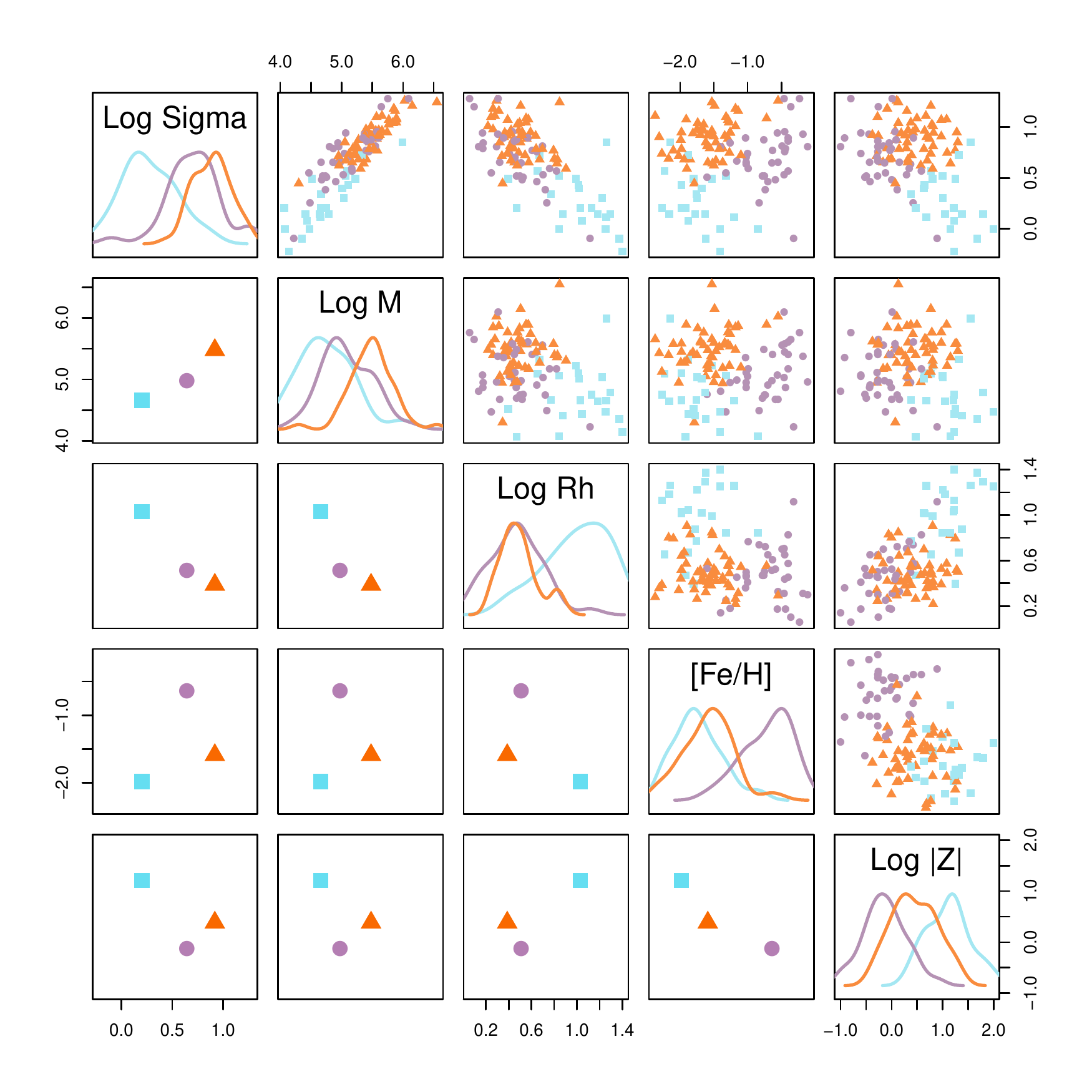}
\caption{Pair plot for all the variables of the dataset. In the top right half of the plot all GCs are shown, color and shape-coded to the cluster they are assigned to as in Fig.~\ref{PC1PC2}. On the bottom left half only the group medoids are shown. On the diagonal the distribution of each quantity is shown for the three groups, obtained by kernel density estimation (scaled so the maxima are the same for the three groups) and color coded accordingly. \label{pairs}}
\end{figure}

We find that our GC sample is naturally divided in either $2$ or (slightly preferred) $3$ clusters. To show this, we plot the average silhouette widths -listed also in Tab.~\ref{silhou}- as a function of the number of clusters $2 \le k \le 10$ obtained by PAM in Fig.~\ref{Optimal}. The highest value of $0.290 \pm 0.015$ corresponds to the best clustering structure on the dataset, and is obtained for $k=3$.
However, $k=2$ returns $0.288 \pm 0.014$, a very similar value. The quoted error bars are one standard deviation of the mean, i.e. $\sigma_{\mathtt{sample}} /  \sqrt{N - 1}$. We adopted these error bars because the average silhouette width is actually an average of the silhouette widths of all points in our sample, so we expect its distribution under repeated sampling to follow the central limit theorem.
With the adopted error bars we see that the average silhouette width for $k=2$ is not significantly different from the $k=3$ one.
In the following we discuss the $k=3$ case and compare some aspects of the resulting clustering to the $k=2$ case.
Tab.~\ref{clusters} shows a summary of the three clusters we obtained. The names we assign to each cluster depend on the properties of the respective medoid, listed in Tab.~\ref{medoids}: for example, the \emph{disk} group medoid, NGC $6352$, has high metallicity [Fe$/$H] $= -0.64$ and distance from the Galactic plane of only $724$pc. A comprehensive view of the properties of medoids is afforded by Fig.~\ref{medoidspider}. In that figure and in the following we assign colors to the three groups as follows: violet for the disk group, orange for the inner halo, and light blue for the outer halo.
Fig.~\ref{pairs} shows the distributions of each variable estimated with Kernel Density Estimation (diagonal panel) for the three groups, together with a view of all the GCs, shape- and color-coded by group, in all planes obtained by plotting two out of the five variables (top-right panels above the diagonal), and a view of the medoids on the same planes (bottom left panels).

\begin{table}
\caption{Silhouette width as a function of the number of clusters.\label{silhou}}
\begin{tabular}{lllll}
\hline
\hline
$k$ & mean s. w. & 1st quart. s. w.& median s. w. & 3rd quart. s. w.\\
\hline
$2$ & $0.288 \pm 0.014$ & $0.192$ & $0.315$ & $0.400$\\
$3$ & $0.290 \pm 0.015$ & $0.149$ & $0.333$ & $0.428$\\
$4$ & $0.255 \pm 0.014$ & $0.133$ & $0.283$ & $0.371$\\
$5$ & $0.246 \pm 0.015$ & $0.111$ & $0.269$ & $0.359$\\
$6$ & $0.221 \pm 0.015$ & $0.106$ & $0.239$ & $0.332$\\
$7$ & $0.225 \pm 0.014$ & $0.107$ & $0.239$ & $0.327$\\
$8$ & $0.228 \pm 0.014$ & $0.137$ & $0.228$ & $0.335$\\
$9$ & $0.225 \pm 0.013$ & $0.129$ &$0.222$ & $0.328$\\ 
$10$ & $0.216 \pm 0.015$ & $0.111$ & $0.221$ & $0.327$\\
\end{tabular}
\end{table}

\begin{table}
\caption{Summary of the clusters obtained for $k=3$. Col.~$1$ lists the names assigned to the clusters based on their physical properties, Col.~$2$ the medoid GC name, Col.~$3$ the number of GCs in each cluster, Col.~$4$ the parameter-space distance between the furthest elements of each cluster (its \emph{diameter}), Col.~$5$ the average distance of its members from those of its neighboring cluster.\label{clusters}}
\begin{tabular}{lllll}
\hline
\hline
Cluster & Medoid & No. of elements & Diameter & Separation\\
\hline
Disk & NGC $6352$ & $36$ & $2.87$ & $0.14$\\ 
Inner halo & NGC $5986$ & $52$ & $2.45$ & $0.14$\\ 
Outer halo & NGC $5466$ & $22$ & $2.53$ & $0.32$\\ 
\end{tabular}
\end{table}

\begin{table}
\caption{Properties of the medoids of the clusters obtained for $k=3$. \label{medoids}}
\begin{tabular}{llllll}
\hline
\hline
Medoid &  $\log \sigma_0$ & $\log M$ & $\log R_h$ & [Fe$/$H] & $\log |Z|$\\
 &  (km$/$s) & (M$_\odot$) & (pc) &  & (kpc)\\
\hline
NGC $6352$ & $0.64$  & $5.00$ & $0.51$ & $-0.64$ & $-0.14$ \\ 
NGC $5986$ & $0.92$ &  $5.48$ & $0.39$ & $-1.59$ & $0.38$ \\ 
NGC $5466$ & $0.20$  & $4.66$ & $1.03$  & $-1.98$ & $1.21$\\ 
\end{tabular}
\end{table}

\begin{figure}
\includegraphics[width = 0.95\columnwidth]{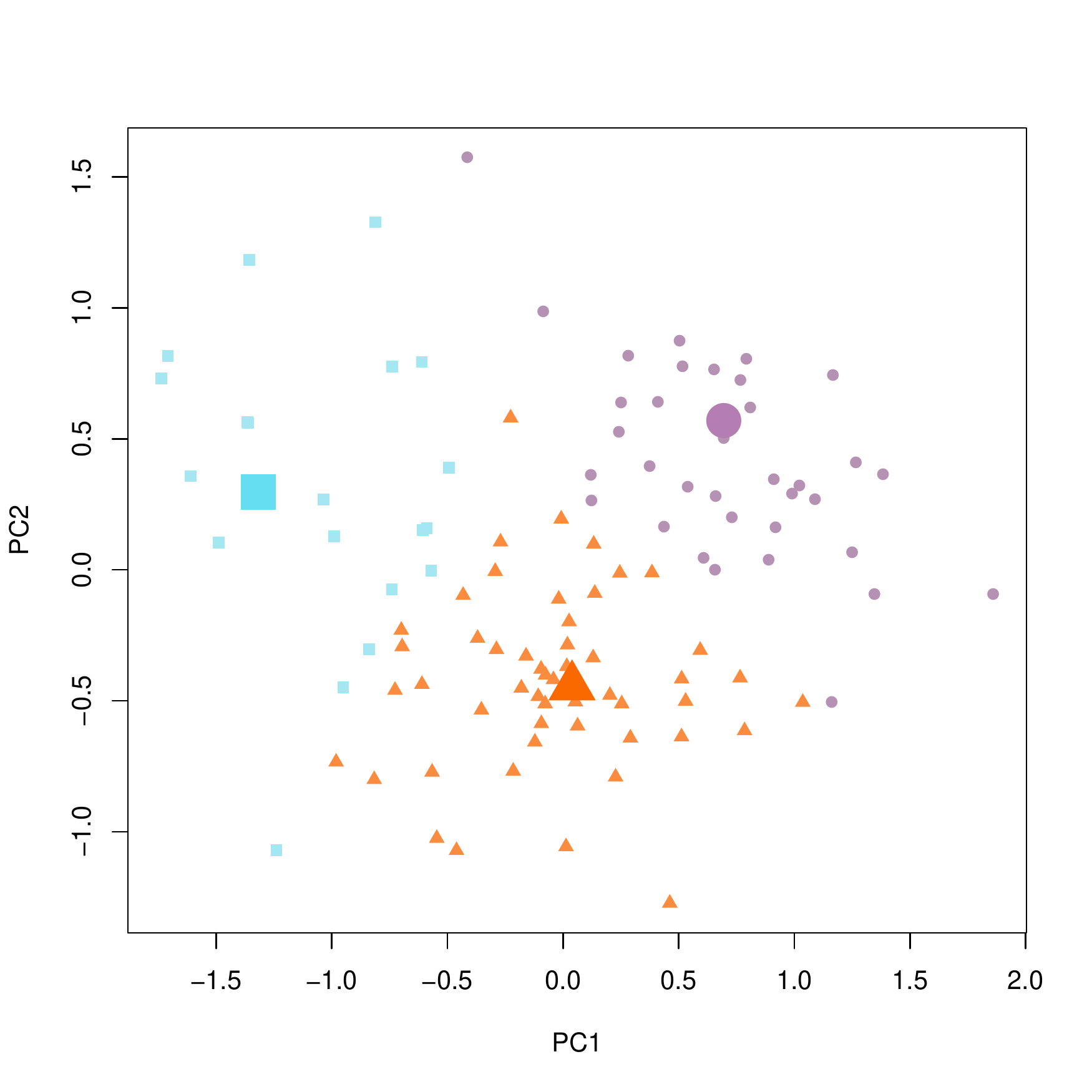}
\caption{Plot of the three groups obtained by PAM for $k=3$ (disk $=$ violet circles, inner halo $=$ orange triangles, and outer halo $=$ light blue squares) in the first two principal components. The respective medoids (NGC $6352$, NGC $5986$, and NGC $5466$) are plotted as bigger and slightly more color-saturated symbols.\label{PC1PC2}}
\end{figure}

Fig.~\ref{PC1PC2} shows the three clusters plotted in the plane of the first two Principal Components (PCs). These two principal components explain about $80\%$ of the total sample variance.

\subsection{Comparison with the $k=2$ case}
\begin{figure}
\includegraphics[width = 0.95\columnwidth]{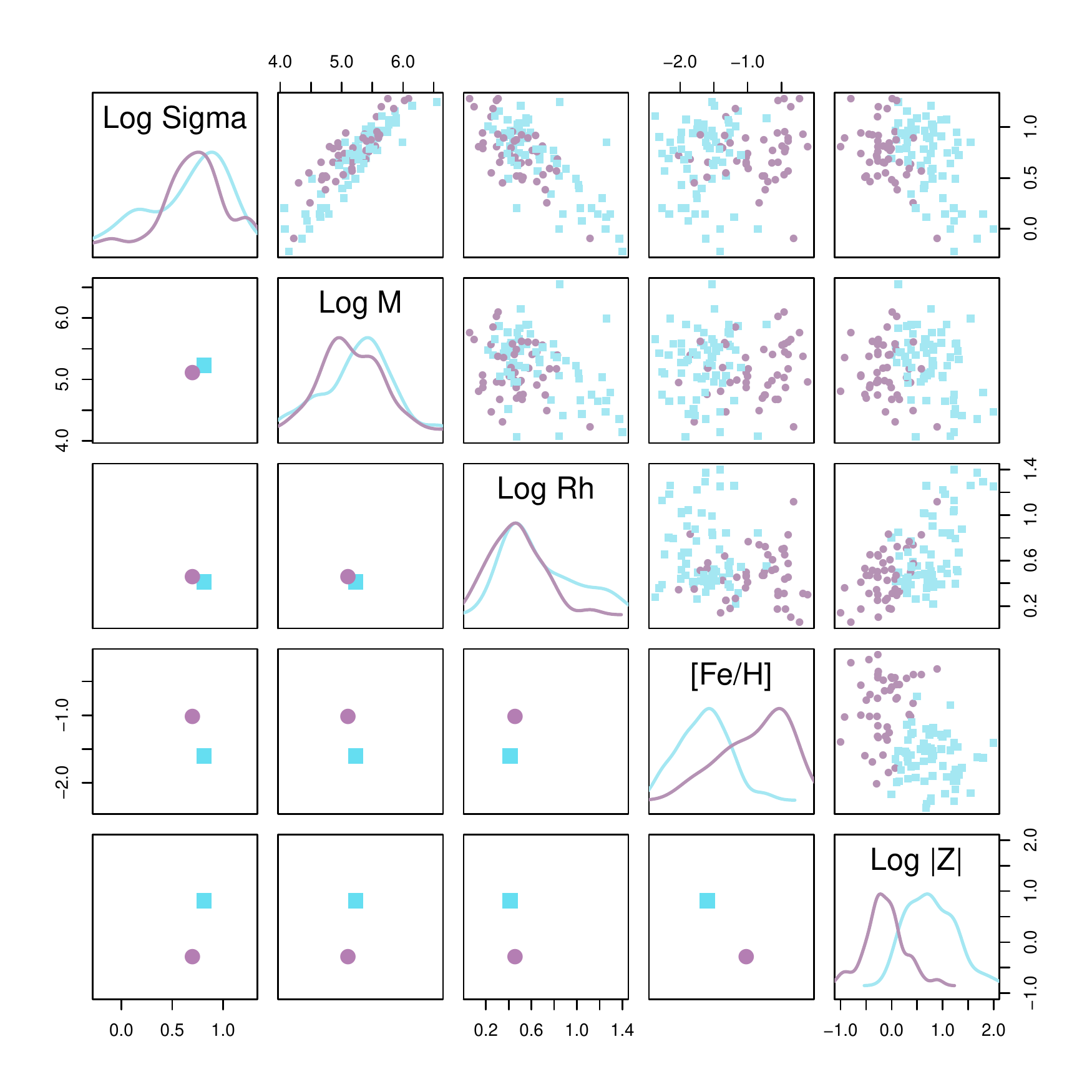}
\caption{Pair plot for all the variables of the dataset. In the top right half of the plot all GCs are shown, color and shape-coded to the cluster they are assigned to as in Fig.~\ref{PC1PC2Only2}. On the bottom left half only the group medoids are shown. On the diagonal the distribution of each quantity is shown for the three groups, obtained by kernel density estimation (scaled so the maxima are the same for the two groups; notice how now the outer halo group of Fig.~\ref{pairs} seems to have disappeared; this is due to the fact that the inner halo group with which it is merged is much more numerous even though this scaling made it not readily apparent in Fig.~\ref{pairs}) and color coded accordingly. \label{pairsOnly2}}
\end{figure}

\begin{figure}
\includegraphics[width = 0.95\columnwidth]{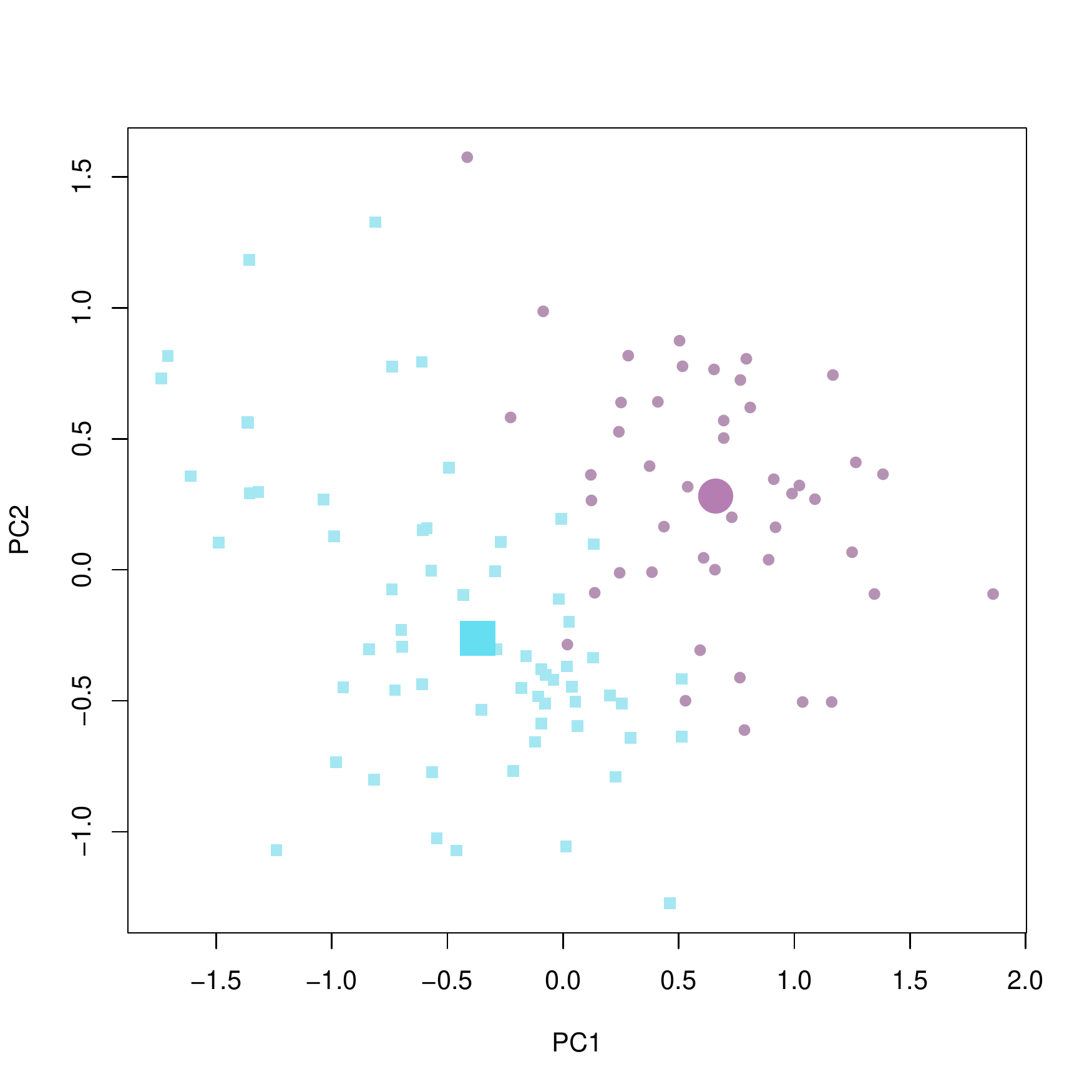}
\caption{Plot of the two clusters obtained by PAM for $k=2$ (disk $=$ violet circles, halo $=$ light blue squares) in the first two principal components. The medoids (now NGC $1904$ and NGC $6712$ respectively) are plotted as bigger and slightly more color-saturated symbols. Comparing with Fig.~\ref{PC1PC2} we see that the inner halo cluster was split among the halo and disk clusters, while disk clusters and outer halo clusters were not affected. \label{PC1PC2Only2}}
\end{figure}

Repeating our analysis with $k=2$ we obtain essentially a halo/disk dichotomy. GCs that were assigned to the inner halo group with $k=3$ are now split among the two disk and halo clusters, while outer halo GCs are all now in the halo group and disk GCs stay in the disk group. This can be appreciated visually in Fig.~\ref{pairsOnly2} and especially in Fig.~\ref{PC1PC2Only2}. In both figures we color coded the halo group in light blue (matching the outer halo group for $k=3$) and the disk group in violet (matching the disk group for $k=3$).

\subsection{Changing the weight of structural parameters}
By comparing Fig.~\ref{pairs} with Fig.~\ref{pairsOnly2} we see how the height on the Galactic plane $Z$ and metallicity are sufficient for dividing GCs in two groups (disk/halo) and the further subdivision of the halo group into two (inner halo / outer halo) is due to differences in structural parameters (outer halo clusters have small mass, small velocity dispersion, and large radius). It can also be observed that the medoids of the two groups are very similar in the three structural parameter variables $(\log M, \log \sigma_0, \log R_e)$. To measure the relative importance of the structural parameter variables versus that of the chemical/orbital variables  $([Fe/H], \log | Z |)$ in determining the preferred number of clusters, we introduced a transformation of the coordinates as follows:
\begin{equation}
\begin{split}
(\log M, \log \sigma_0, \log R_e, [Fe/H], \log | Z |) \to \\
\to (p \log M, p \log \sigma_0, p \log R_e, [Fe/H], \log | Z |)
\end{split}
\end{equation}
and we let the \emph{stretching parameter} $p$ vary from $1/3$ to $3$, and re-run PAM obtaining the average silhouette widths for $k=2$ and $k=3$ as a function of $p$. PAM is using Euclidean distances to compute dissimilarities between GCs in the transformed coordinates, so the transformation results in weighing the distances in the $(\log M, \log \sigma_0, \log R_e)$ subspace more (for $p > 1$) or less (for $p < 1$) with respect to those in the $([Fe/H], \log | Z |)$ subspace for the purposes of clustering. Technically this corresponds to using PAM with a distance induced by the Mahalanobis norm
\begin{equation}
||\mathbf{v}|| = \mathbf{v}^T \mathbf{A} \mathbf{v}
\end{equation}
where $\mathbf{v}$ is a point in the $(\log M, \log \sigma_0, \log R_e, [Fe/H], \log | Z |)$ space and $\mathbf{A}$ is a diagonal matrix in the form
\begin{equation}
\mathbf{A} = 
\begin{bmatrix}
    p      & 0 & 0 & 0 & 0 \\
    0       & p & 0 & 0 & 0 \\
   0       & 0 & p & 0 & 0 \\
   0 & 0 & 0 & 1 & 0 \\
   0 & 0 & 0 & 0 & 1\\
\end{bmatrix}
\end{equation}
In the limit of $p \to 0$ we would be clustering only in $([Fe/H], \log | Z |)$ and for $p \to \infty$ only in $(\log M, \log \sigma_0, \log R_e)$. 

\begin{figure}
\includegraphics[width = 0.95\columnwidth]{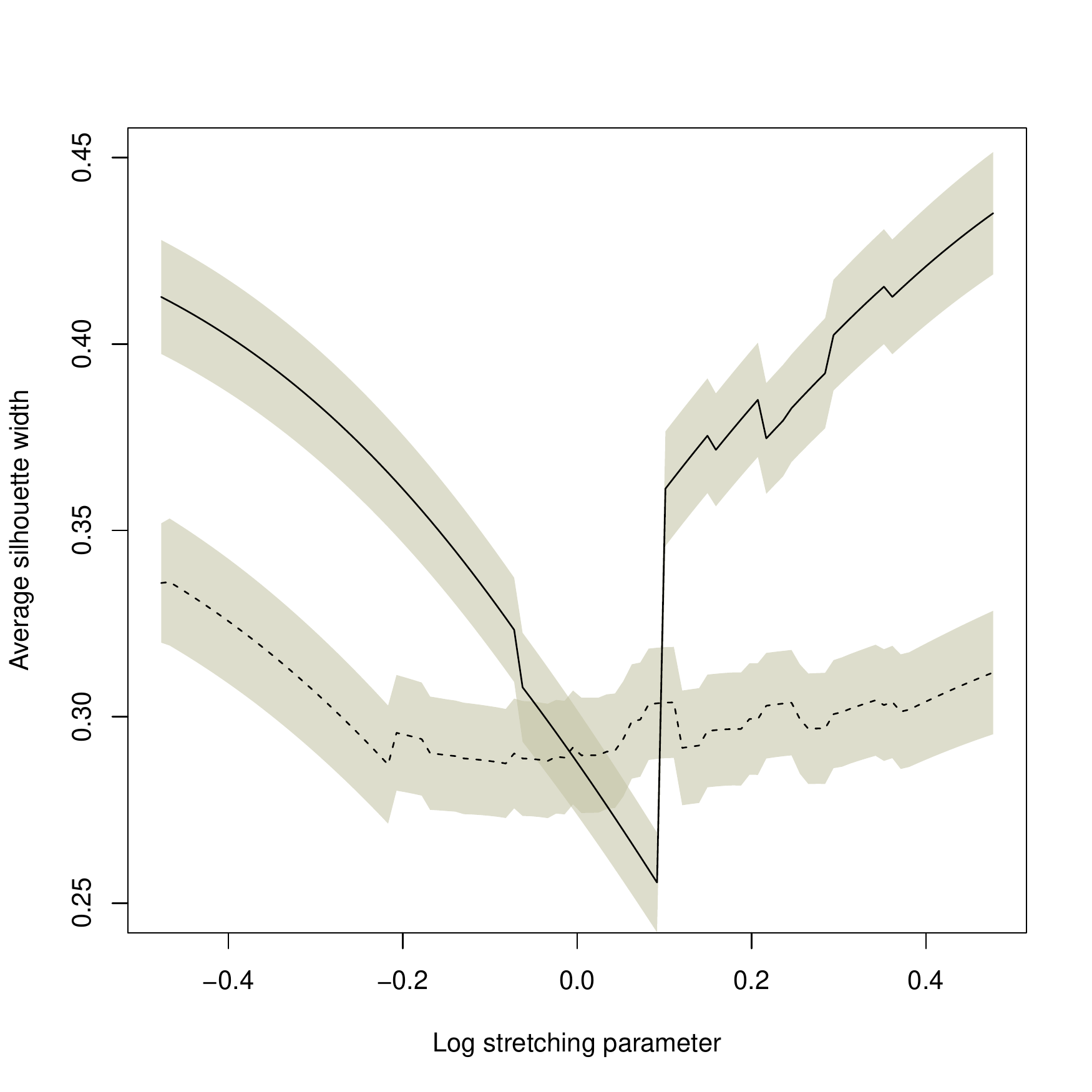}
\caption{Average silhouette width of $k=2$ clustering structure (solid line) and $k = 3$ (dashed line) as a function of the stretching parameter $p$. One-sigma errors on the average are represented by gray shaded areas. Values of $p>1$ correspond to weighing structural parameters $\log M$, $\log \sigma_0$, and $\log R_e$ more than chemical/orbital parameters [Fe$/$H] and $\log | Z |$, and $p<1$ values do the reverse.\label{stretching}}
\end{figure}

Fig.~\ref{stretching} shows the results of clustering in the transformed space: the average silhouette width for $k = 2$ (solid line) is higher than that for $k = 3$ (dashed line) for all values of $p$ (so two clusters are preferred over three) except in a region around $p = 1$ where the average silhouette width for $k = 2$ drops below that for $k = 3$. This can be interpreted as evidence that three clusters are preferred only if both structural and chemical/orbital parameters are considered.

\subsection{Silhouettes}
\begin{figure}
\includegraphics[width = 0.95\columnwidth]{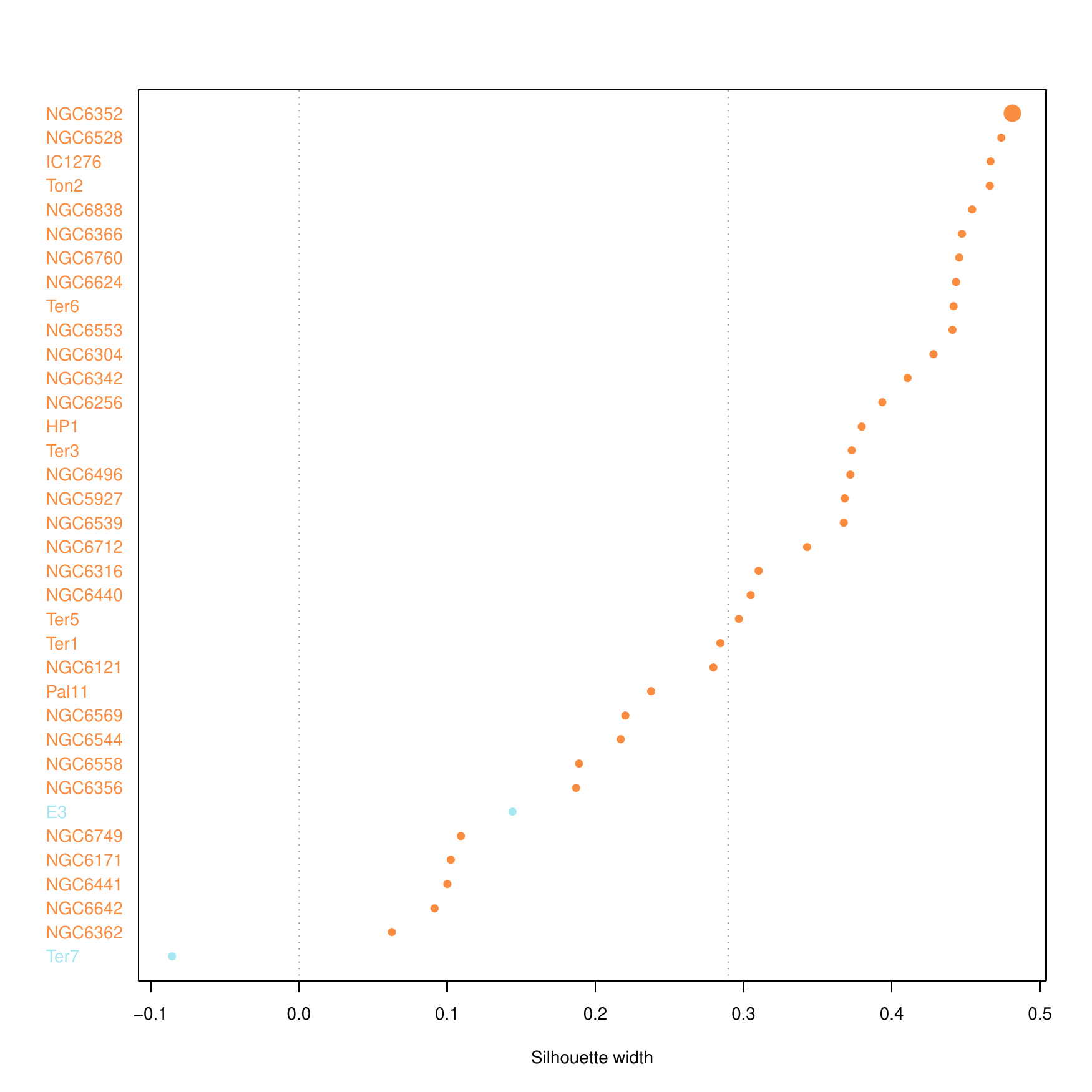}
\caption{Silhouettes widths for each member of the disk group. The color coding represents the neighboring group (light blue, outer halo group; orange, inner halo group) to which each GC is most similar, i.e. the second best group; predictably most disk GCs are neighbors of the inner halo group. The vertical dotted lines correspond to $0$ and to the average silhouette width for the whole dataset. GCs that lie to the left of the leftmost vertical dotted line have negative silhouettes, which suggest that they are more similar to a neighboring group rather than to the group that they have been assigned to. The bigger symbol is the group medoid, who needs not have the highest value of silhouette width.\label{sil2}}
\end{figure}

\begin{figure}
\includegraphics[width = 0.95\columnwidth]{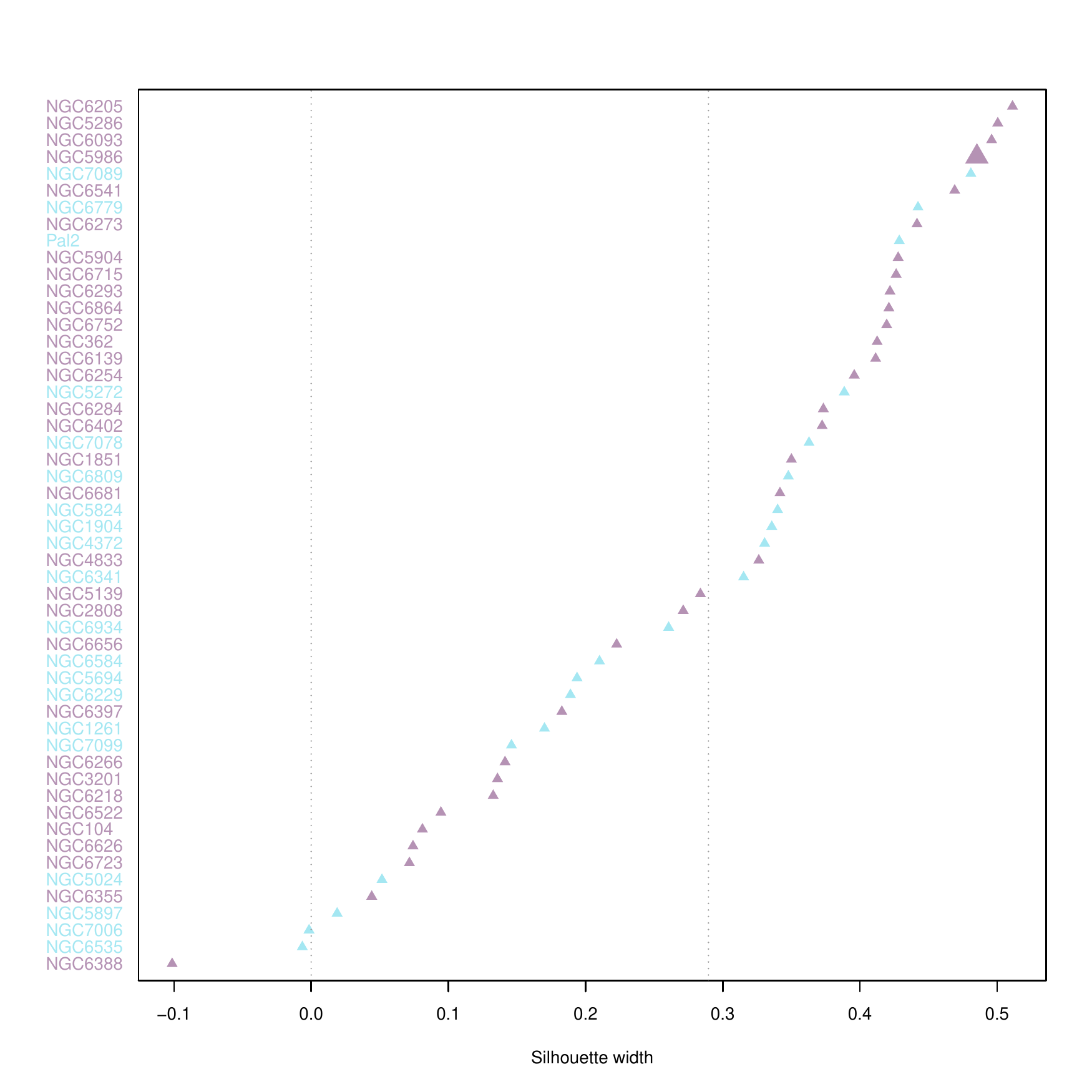}
\caption{Silhouettes widths for each member of the inner halo group. The color coding represents the neighboring group (light blue, outer halo group; violet, disk group) to which each GC is most similar, i.e. the second best group; some inner halo GCs neighbor on the disk group, some on the outer halo group. The vertical lines correspond to $0$ and to the average silhouette width for the whole dataset. GCs that lie to the left of the leftmost vertical dotted line have negative silhouettes, which suggest that they are more similar to a neighboring group rather than to the group that they have been assigned to. The bigger symbol is the group medoid, who needs not have the highest value of silhouette width.\label{sil3}}
\end{figure}

\begin{figure}
\includegraphics[width = 0.95\columnwidth]{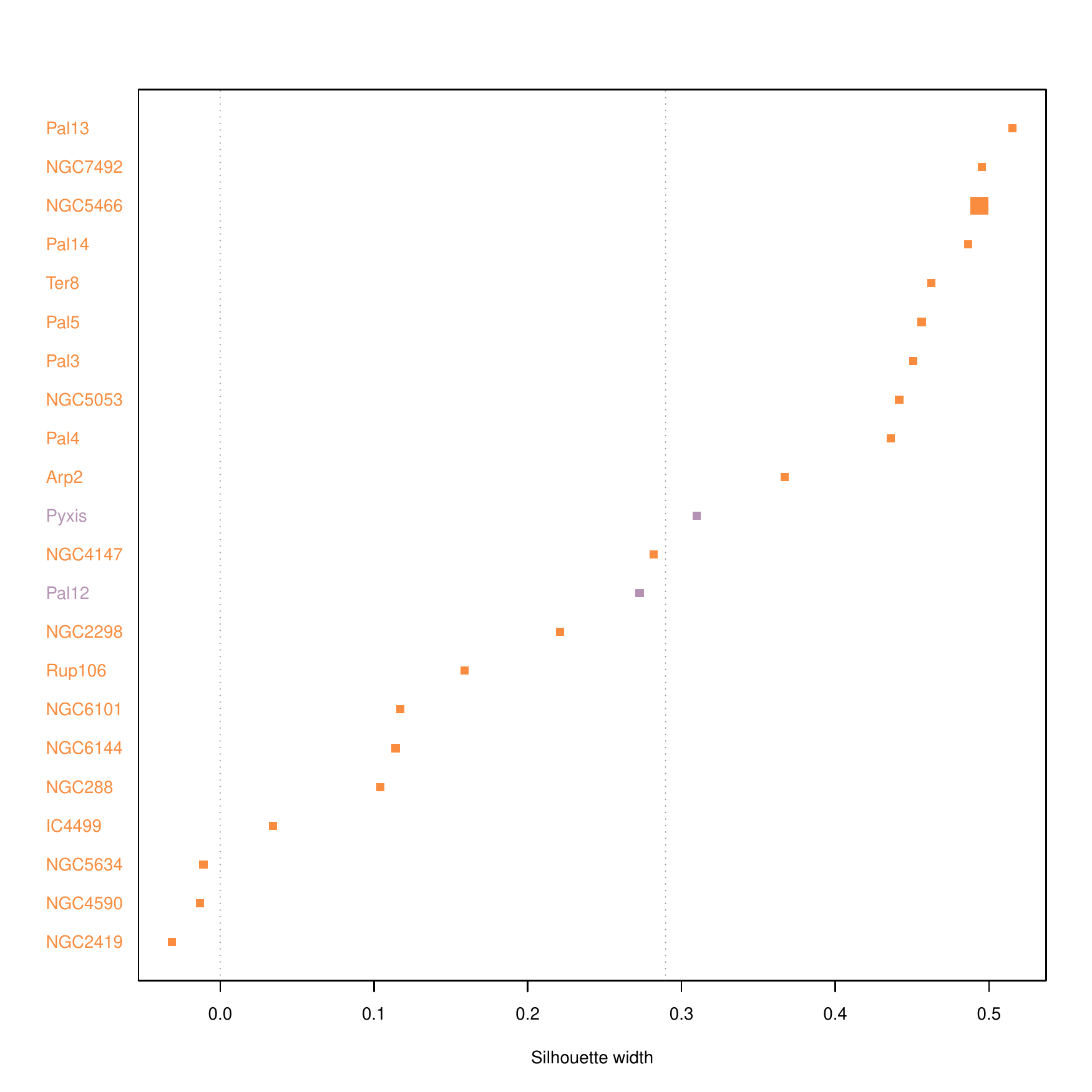}
\caption{Silhouette widths for each member of the outer halo group. The color coding represents the neighboring group (violet, disk group; orange, inner halo group) to which each GC is most similar i.e. the second best group; as for the disk group, most outer halo GCs have the inner halo group as neighbor. The vertical lines correspond to $0$ and to the average silhouette width for the whole dataset. GCs that lie to the left of the leftmost vertical dotted line have negative silhouettes, which suggest that they are more similar to a neighboring group rather than to the group that they have been assigned to. The bigger symbol is the group medoid, who needs not have the highest value of silhouette width.\label{sil1}}
\end{figure}

In Fig.~\ref{sil2}, \ref{sil3}, and \ref{sil1} we show the silhouettes of the disk, inner halo and outer halo group respectively. In each figure GCs are arranged by decreasing silhouette width and color coded according to their neighboring cluster. Low silhouette width GCs are in-between groups.

\subsection{A posteriori analysis of other physical parameters across the groups}
\begin{figure}
\includegraphics[width = 0.95\columnwidth]{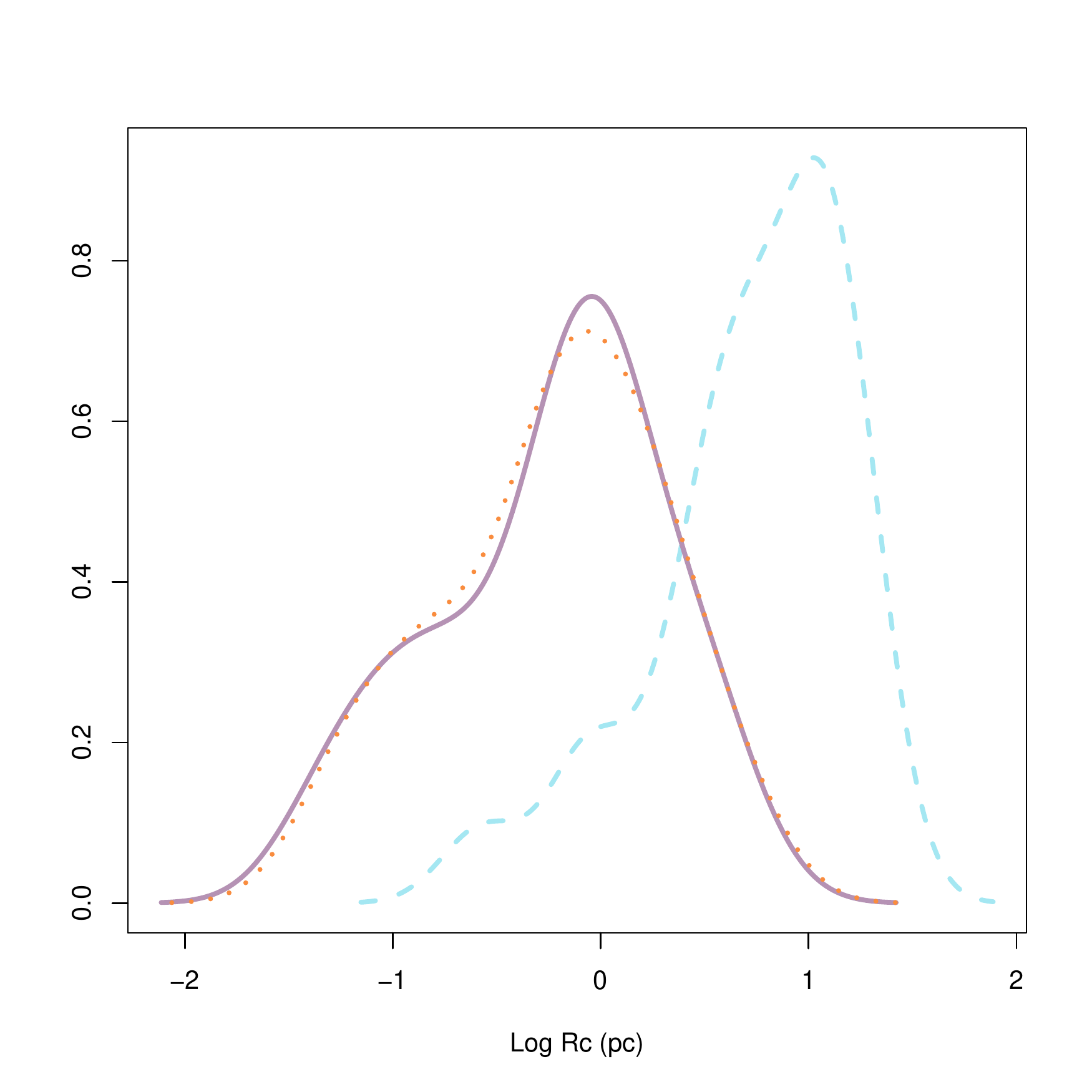}
\caption{Log core radius distributions estimated with kernel density estimation for the three groups: disk (solid purple line), inner halo (dotted orange line), and outer halo (dashed light blue line). \label{comparecoreradius}}
\end{figure}

\begin{figure}
\includegraphics[width = 0.95\columnwidth]{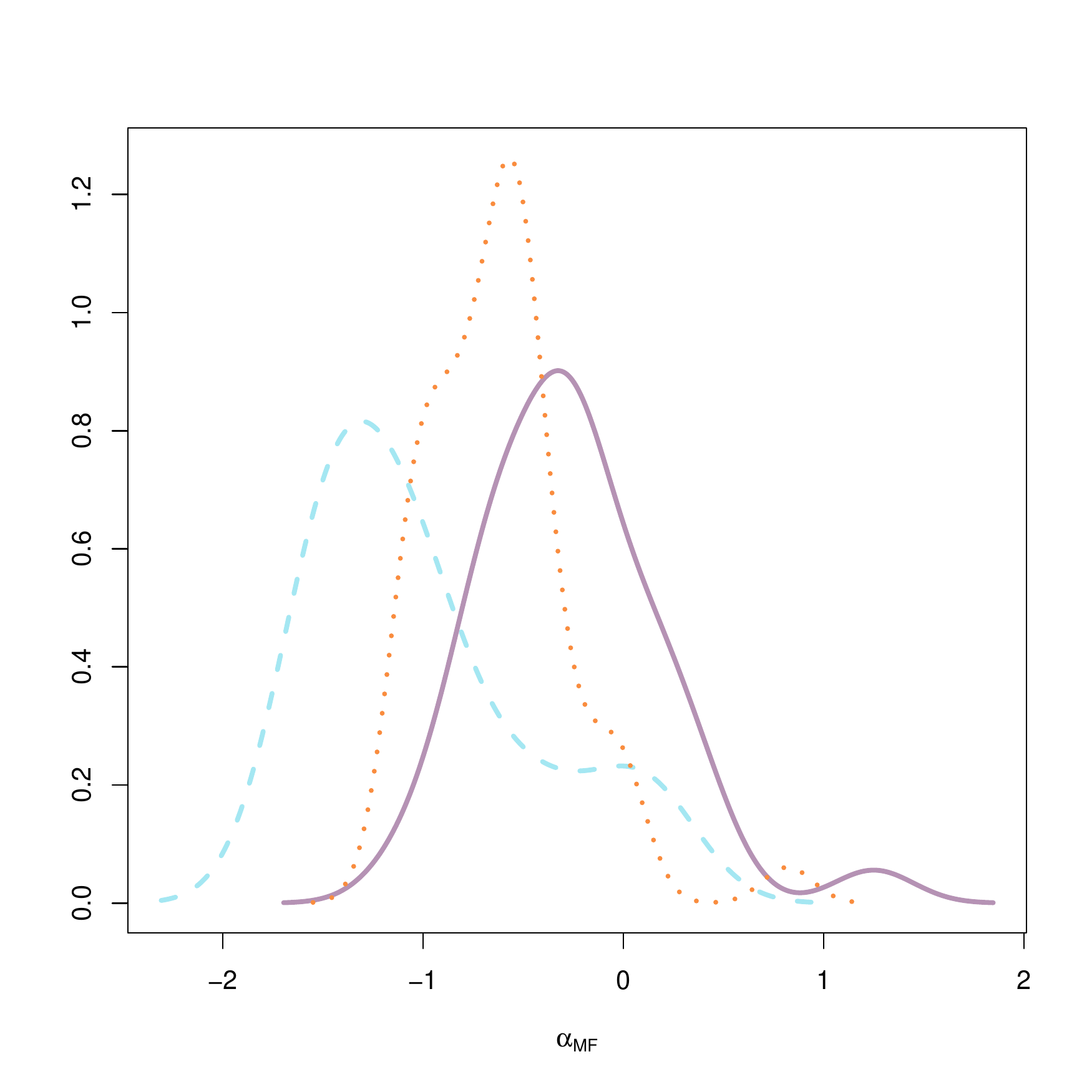}
\caption{Mass function slope distributions estimated with kernel density estimation for the three groups: disk (solid purple line), inner halo (dotted orange line), and outer halo (dashed light blue line). \label{compareMFslope}}
\end{figure}

\begin{figure}
\includegraphics[width = 0.95\columnwidth]{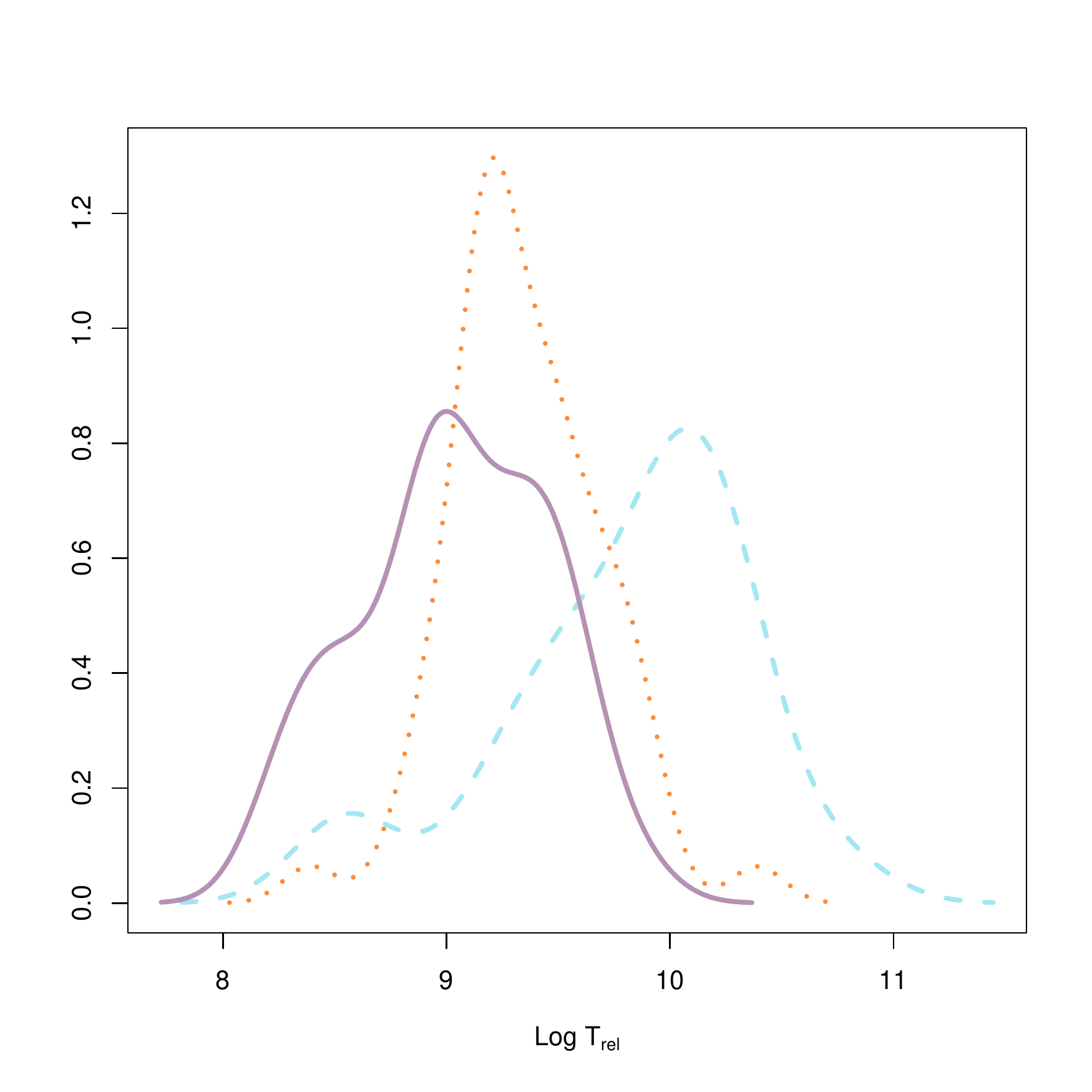}
\caption{Log half-mass relaxation time distributions estimated with kernel density estimation for the three groups: disk (solid purple line), inner halo (dotted orange line), and outer halo (dashed light blue line). \label{comparerelaxationtime}}
\end{figure}

In Fig.~\ref{comparecoreradius} we compare the core radii distributions from \cite{2018MNRAS.478.1520B} for the three groups. The plots are obtained by kernel density estimation. Disk and inner halo clusters have very similar distributions of log core radius; any differences are not significant according to a Kolmogorov-Smirnov (KS) test. On the other hand the top panel shows that outer halo clusters tend to have much larger cores with respect to the two other groups, and this difference is significant to $2 \times 10^{-8}$ according to a KS test. Larger cores in the outer halo are likely due to the effect of the Galactic tidal field, which is stronger for inner halo and disk GCs.
Fig.~\ref{compareMFslope} compares the global, present day stellar mass function slope also from \cite{2018MNRAS.478.1520B}. The distributions are all significantly different from each other, with a KS test p-value $< 2 \times 10^{-4}$. Outer halo clusters have mostly negative values of the mass function slope, most likely because they are less dynamically evolved (see later) due to their lower average density, and consequently did not yet lose low-mass stars through dynamical evaporation. Additionally, and perhaps more importantly, outer halo clusters do not suffer from tidal stripping by the Galaxy, which is also removing preferentially low-mass stars. Disk clusters on the other hand have very depleted mass functions, often with positive slopes. Interestingly, inner halo clusters behave in an intermediate way between outer halo and disk clusters. This very clear cut subdivision by mass function slope mirrors that in Fig.~\ref{comparerelaxationtime} by relaxation time. This is perhaps unsurprising because relaxation time is the timescale for the evolution of a GC mass function through collisional processes, but also because \cite{2018MNRAS.478.1520B} obtained mass-function slopes for GCs where a direct measurement was missing by using an empirical relation with relaxation time \citep[][]{2017MNRAS.471.3668S}.

\subsection{Differences in age among and across the groups}
\begin{figure}
\includegraphics[width = 0.95\columnwidth]{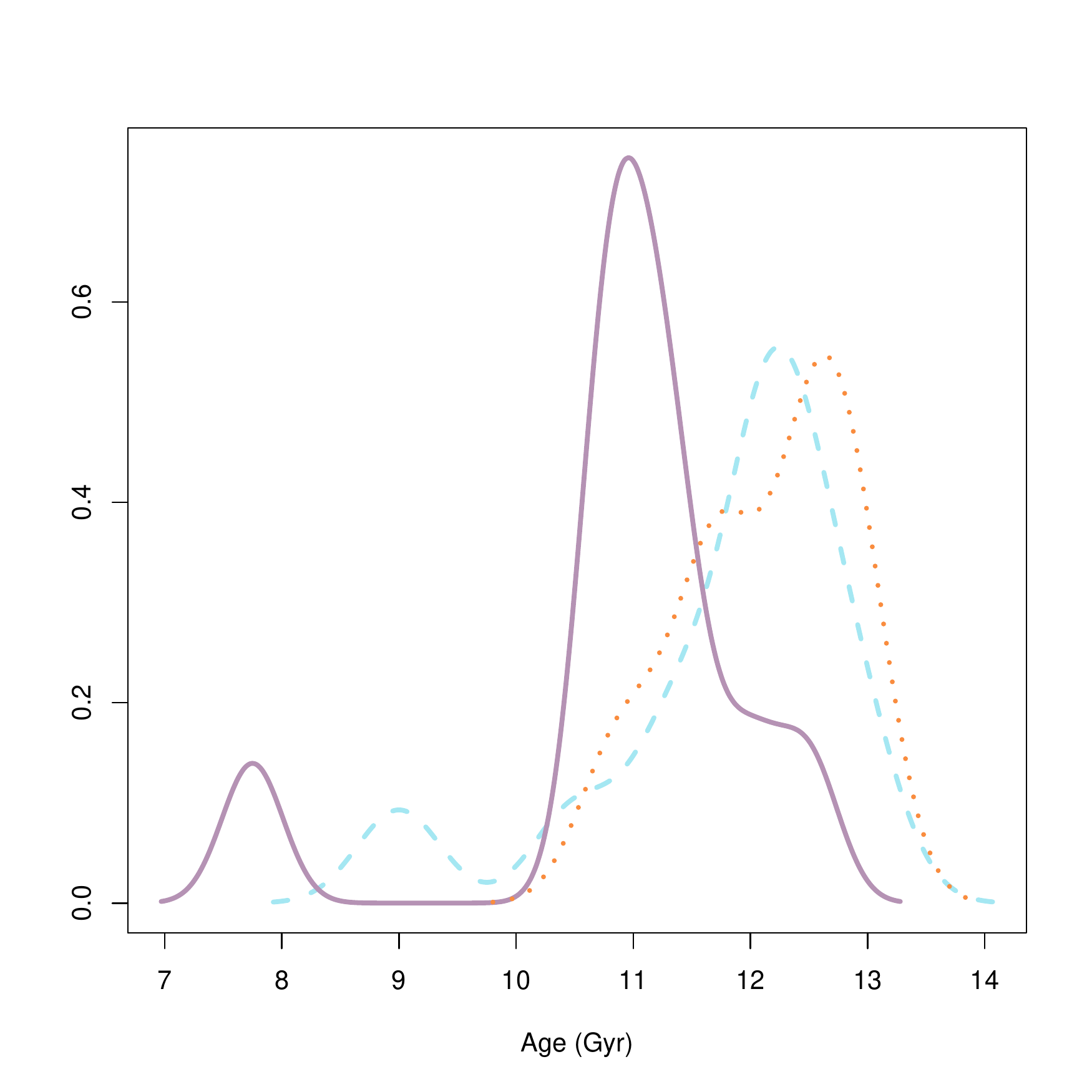}
\caption{Age distributions estimated with kernel density estimation for the three groups: disk (solid purple line), inner halo (dotted orange line), and outer halo (dashed light blue line). The bumps at about $8$ and $9$ Gyr are due to Terzan $7$ and Palomar $12$, which are outliers in the distributions of their respective groups (disk and outer halo).\label{compareAge}}
\end{figure}

\begin{figure}
\includegraphics[width = 0.95\columnwidth]{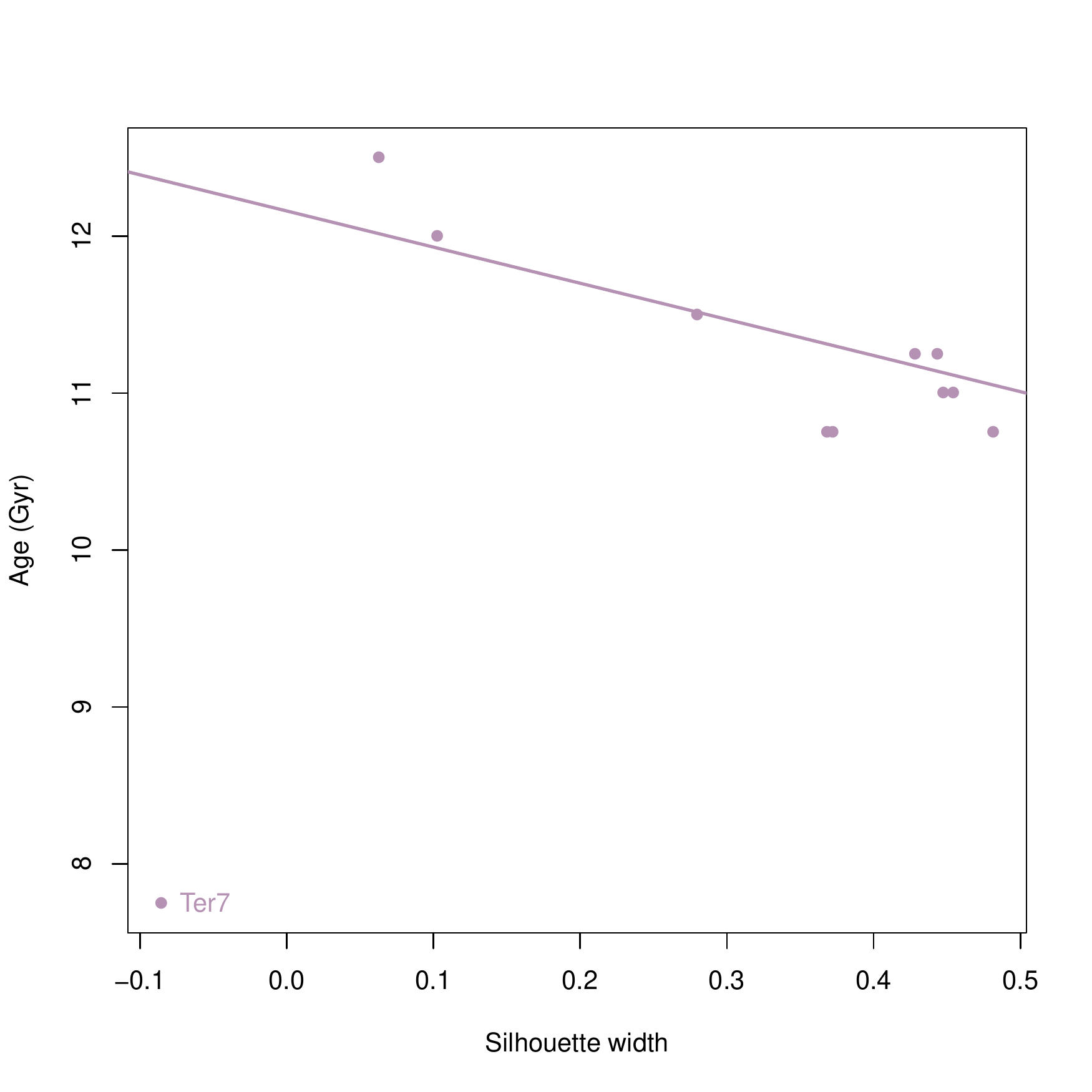}
\caption{Correlation between silhouette width and age for disk GCs. GCs that are more strongly associated with the disk group (have a higher silhouette width) are younger, despite the fact that no information on age was used to determine the clustering and the relevant silhouette widths. The line is a robust linear fit, and one outlier (Terzan $7$) is labeled.\label{SilhouetteCorrelation2}}
\end{figure}

\begin{figure}
\includegraphics[width = 0.95\columnwidth]{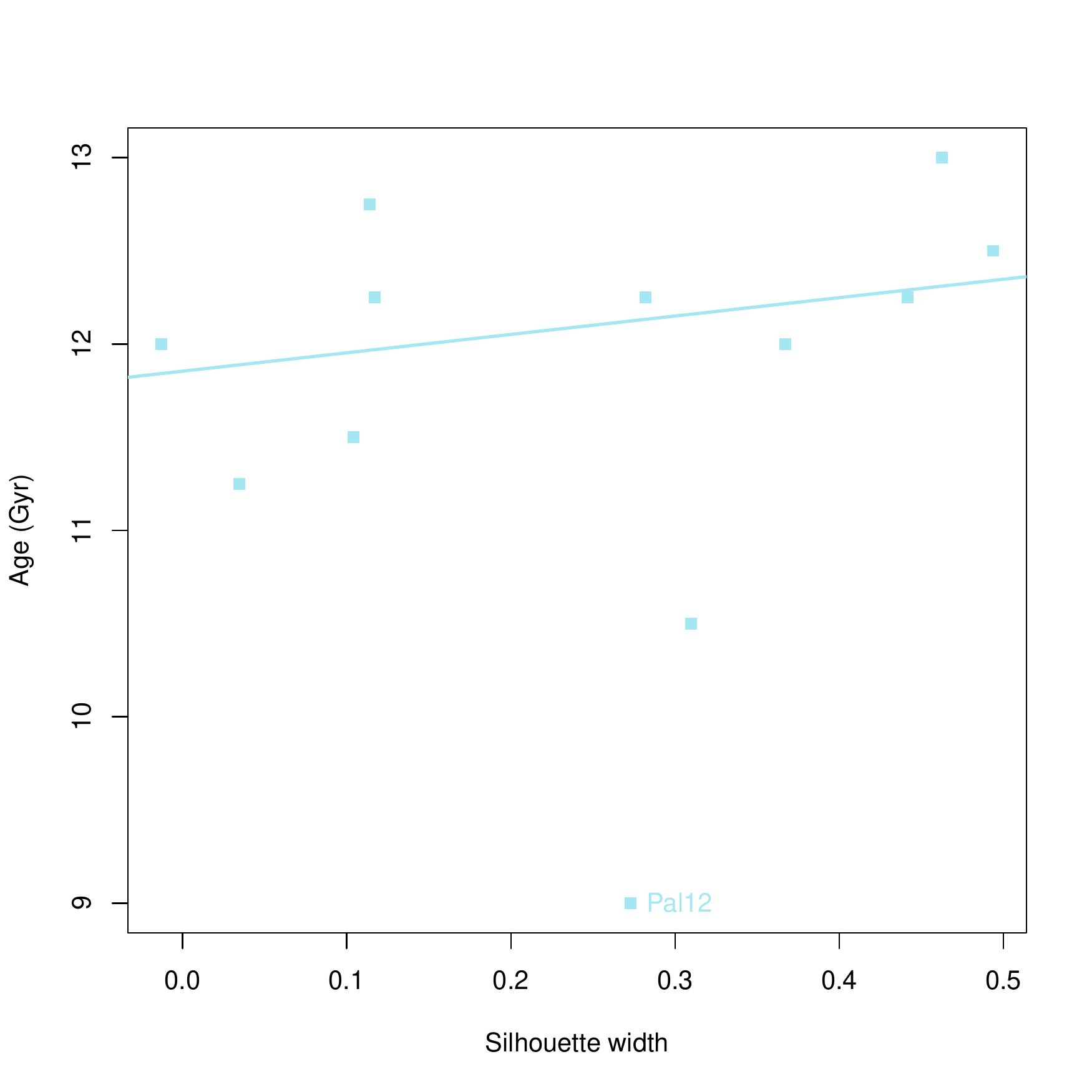}
\caption{Correlation between silhouette width and age for outer halo GCs. GCs that are more strongly associated with the outer halo group (have a higher silhouette width) are older, despite the fact that no information on age was used to determine the clustering and the relevant silhouette widths. The line is a robust linear fit, and one outlier (Palomar $12$) is labeled.\label{SilhouetteCorrelation1}}
\end{figure}

Another interesting comparison, which also shows the value of silhouette widths as a measure of how strongly a given GC is associated with its assigned group, is with GC age. We obtained GC ages from \cite{2018arXiv181104798R} Tab.~1, which has an overlap of $54$ GCs with our main sample.
Fig.~\ref{compareAge} shows, as before, the age distribution across the three groups. Predictably, disk GCs are younger, and inner halo and outer halo GCs are older. However, a very interesting correlation is shown in Fig.~\ref{SilhouetteCorrelation2} and Fig.~\ref{SilhouetteCorrelation1}.
In the first figure, as the silhouette width increases, meaning that a given GC is more and more confidently classified as a disk GC, age decreases. The more a GC is disk-like, the younger it is. In the second figure, the reverse happens for outer halo GCs: increasing silhouette widths correspond to higher ages. This is remarkable as the disk / inner halo / outer halo groups were not obtained using age information, and goes a long way to show the usefulness of silhouette width as measure of the quality of clustering.
In both cases only two clusters do not respect the silhouette width versus age relation: Palomar $12$ for the halo group, and Terzan $7$ for the disk group. Both Palomar $12$ and Terzan $7$ are peculiar not only for strongly differing in age from their assigned group, but also in that they are among the few members of the outer halo and disk groups respectively whose neighboring cluster (their second best choice for classification) is not the inner halo.
Additionally Terzan $7$ has the lowest silhouette width for the disk group, suggesting that it may be better off assigned to the neighboring outer halo group instead (Palomar $12$ is firmly classified as an outer halo cluster instead).

\subsection{Fraction of first generation stars}
\begin{figure}
\includegraphics[width = 0.95\columnwidth]{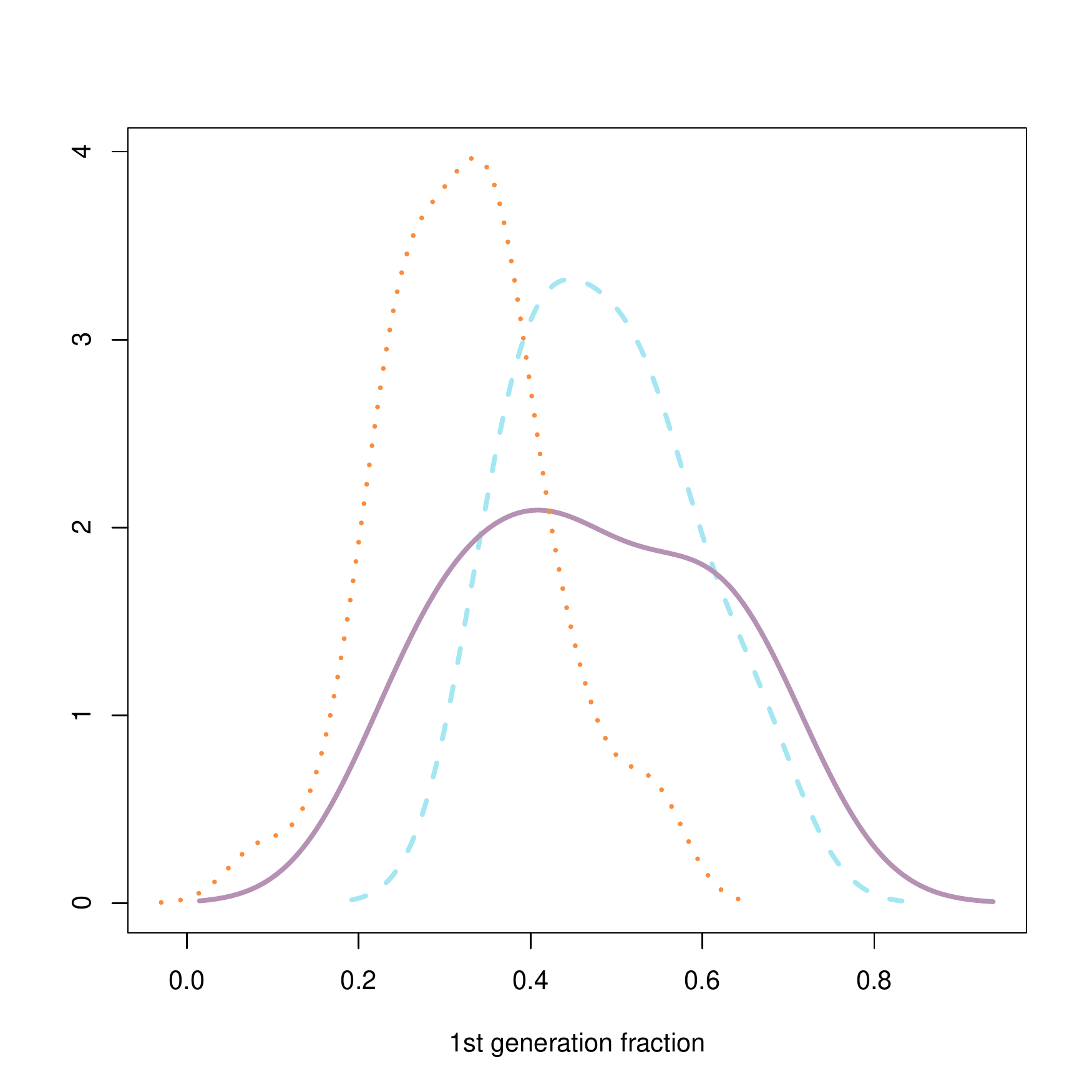}
\caption{First generation star fraction distributions estimated with kernel density estimation for the three groups: disk (solid purple line), inner halo (dotted orange line), and outer halo (dashed light blue line). \label{firstgenfrac}}
\end{figure}

We used estimates of the fraction of first generation (FG) stars by \cite{2017MNRAS.464.3636M} to check whether there are any systematic differences between the three groups in FG star abundance. The overlap between our table and \cite{2017MNRAS.464.3636M} is such that only $8$ GCs with measured FG fraction are contained in the disk group, $35$ in the inner halo group, and  only $7$ in the outer halo group. With these numbers in mind, we run Kolmogorov-Smirnov tests to measure the significance of any differences in FG star fraction, and found that inner- and outer-halo GCs have significantly different FG fraction ($p = 0.001$), inner-halo and disk marginally significantly ($p = 0.04$), while the differences between outer-halo and disk are not significant ($p=0.91$) due to the low number of GCs with a measured FG fraction in each of these groups.
Fig.~\ref{firstgenfrac} clearly shows the difference between inner-halo GCs and the GCs of the two other groups, with inner-halo GCs having in general a lower FG fraction.
FG fraction is the only quantity among those we considered that breaks the pattern where inner-halo GCs are usually in-between disk and outer-halo GCs. This can be interpreted to mean that the physical phenomena driving the formation of multiple stellar generations in GCs are not affected by the disk/inner-halo/outer-halo nature of the host GC, so they are probably related to internal dynamics rather than to the Galactic environment.

\subsection{Comparison with previous results}	
\label{comparison}	
We compare our results with those by F09. Our samples have $46$ GC in common (their sample contains $54$ GCs, only $8$ of which are not part of our sample). Their results are similar to ours, i.e. they obtain three groups which they interpret in terms of disk, inner halo, and outer halo, but both their clustering approach (based on cladistics methods) and their choice of variables (relative ages, metallicity, absolute V magnitude, and maximum effective temperature on the HB) are different from ours. Notwithstanding these differences, our clustering structure is quite compatible with theirs on the overlap of our two samples: less than $20\%$ of GCs are assigned to different groups in the two studies. Furthermore, there are exactly zero GCs that are part of the disk group for us and the halo group for them, and vice versa. 

\begin{table}
\caption{Confusion matrix of our classification against F09 for the shared GCs. Rows correspond to F09 groups and columns to our groups. The numbers in each cell count the number of GCs assigned to the relevant groups, for example the number $3$ in the last row and last column means that three GCs were assigned to the outer halo class by us and by F09 as well. \label{confusioncharbonnel}}
\begin{tabular}{llllll}
\hline
\hline
 & & Our groups & & & \\
 &  & &disk  & halo &\\
  & & & &inner & outer\\
F09 groups &  disk  & &   $11$ &    $1$ &   $0$ \\
 & halo & inner &   $1$   & $23$  &   $0$\\
 & &outer   & $0$  &   $7$   &  $3$\\
\end{tabular}
\end{table}

A detailed comparison of our and F09 classification is presented in Tab.~\ref{confusioncharbonnel}. Off-diagonal numbers represent the count of mismatched GCs between our groups and F09 groups. These are generally low or even zero, except for the case of $7$ GCs classified as inner halo by us and as outer halo by F09. In other terms, our results and F09 share a clear separation between disk and halo GCs, while the inner/outer halo divide is more blurred. This is expected based on what we saw before, i.e. that three groups are just slightly preferred to two groups, based on the average silhouette widths we computed. We also find that the silhouette widths of GCs that are assigned to a different group by us and F09 have a lower average ($0.18$) with respect to our whole sample ($0.29$). This suggests that our classification and F09 differ for clusters that are not fitting well in our classification in the first place.

\section{Conclusions}
It has long been known that GCs are divided in either two (disk / halo) or three (disk / inner halo / outer halo) families, based on a combination of chemical and orbital parameters \citep{1985ApJ...293..424Z, 1993ASPC...48...38Z}. These groupings of GCs were based on simple prescriptions (e.g. a threshold in metallicity) and affected by an element of subjectivity. We attempt to improve on this by using unsupervised machine learning techniques to obtain a clustering of GCs for which the following questions can be readily answered in a quantitative way: a) what is the optimal number of groups? b) how strongly is a given GC associated to its parent group? Previous attempts at obtaining such a subdivision using automated techniques on GC parameters were affected either by a questionable choice of the parameter space to work in, or relied on agglomerative clustering methods and straight out did not attempt to answer question a). No previous work attempts to answer question b), to the best of our knowledge.
In this paper we use the partitioning around medoids algorithm \citep[see][]{KR} in the $(\log M, \log \sigma_0, \log R_e, [Fe/H], \log | Z |)$ parameter space to divide GCs into groups, and the silhouettes introduced by \cite{ROUSSEEUW198753} to estimate the optimal number of groups, answering question a), and to quantify the strength of the association of each GC with its assigned group, answering question b).
We find that the optimal number of groups is three, in the sense that this choice maximizes the average silhouette width over the dataset. We show that the three groups can be identified with the disk / inner halo / outer halo groups, provide representative members for each group (medoids) which have the expected properties (e.g. the medoid for disk GCs is metal rich and near to the Galactic plane), show how additional variables not used for the clustering behave as expected (e.g. disk and outer halo GCs have the lowest and highest ages respectively), and finally find a correlation between silhouette widths and ages for disk GCs (GC that are more confidently classified as members of the disk group are younger) and outer halo GCs (GCs that are more confidently classified as members of the outer halo group are older). We also show that forcing two groups instead of three we obtain, as expected, a disk/halo classification, where the GCs previously assigned to the inner halo are split across the halo and disk groups.

\section*{Acknowledgements}
This project has received funding from the European Union's Horizon $2020$
research and innovation programme under the Marie Sk\l{}odowska-Curie grant agreement No. $664931$.
We thank Prof. Michela Mapelli, Dr. Alessandro Ballone, and Prof. Young-Wook Lee for comments and discussion.

\bibliographystyle{mnras}
\bibliography{manuscript}

\begin{thebibliography}{}
\makeatletter
\relax
\def\mn@urlcharsother{\let\do\@makeother \do\$\do\&\do\#\do\^\do\_\do\%\do\~}
\def\mn@doi{\begingroup\mn@urlcharsother \@ifnextchar [ {\mn@doi@}
  {\mn@doi@[]}}
\def\mn@doi@[#1]#2{\def\@tempa{#1}\ifx\@tempa\@empty \href
  {http://dx.doi.org/#2} {doi:#2}\else \href {http://dx.doi.org/#2} {#1}\fi
  \endgroup}
\def\mn@eprint#1#2{\mn@eprint@#1:#2::\@nil}
\def\mn@eprint@arXiv#1{\href {http://arxiv.org/abs/#1} {{\tt arXiv:#1}}}
\def\mn@eprint@dblp#1{\href {http://dblp.uni-trier.de/rec/bibtex/#1.xml}
  {dblp:#1}}
\def\mn@eprint@#1:#2:#3:#4\@nil{\def\@tempa {#1}\def\@tempb {#2}\def\@tempc
  {#3}\ifx \@tempc \@empty \let \@tempc \@tempb \let \@tempb \@tempa \fi \ifx
  \@tempb \@empty \def\@tempb {arXiv}\fi \@ifundefined
  {mn@eprint@\@tempb}{\@tempb:\@tempc}{\expandafter \expandafter \csname
  mn@eprint@\@tempb\endcsname \expandafter{\@tempc}}}

\bibitem[\protect\citeauthoryear{{Alessandrini}, {Lanzoni}, {Miocchi}, {Ciotti}
   \& {Ferraro}}{{Alessandrini} et~al.}{2014}]{2014ApJ...795..169A}
{Alessandrini} E.,  {Lanzoni} B.,  {Miocchi} P.,  {Ciotti} L.,   {Ferraro}
  F.~R.,  2014, \mn@doi [\apj] {10.1088/0004-637X/795/2/169}, \href
  {http://adsabs.harvard.edu/abs/2014ApJ...795..169A} {795, 169}

\bibitem[\protect\citeauthoryear{Ball \& Hall}{Ball \&
  Hall}{1965}]{ball1965isodata}
Ball G.~H.,  Hall D.~J.,  1965, Technical report, ISODATA, a novel method of
  data analysis and pattern classification.
Stanford research inst Menlo Park CA

\bibitem[\protect\citeauthoryear{{Baumgardt} \& {Hilker}}{{Baumgardt} \&
  {Hilker}}{2018}]{2018MNRAS.478.1520B}
{Baumgardt} H.,  {Hilker} M.,  2018, \mn@doi [\mnras] {10.1093/mnras/sty1057},
  \href {https://ui.adsabs.harvard.edu/#abs/2018MNRAS.478.1520B} {478, 1520}

\bibitem[\protect\citeauthoryear{{Baumgardt}, {Hilker}, {Sollima}  \&
  {Bellini}}{{Baumgardt} et~al.}{2019}]{2019MNRAS.482.5138B}
{Baumgardt} H.,  {Hilker} M.,  {Sollima} A.,   {Bellini} A.,  2019, \mn@doi
  [\mnras] {10.1093/mnras/sty2997}, \href
  {https://ui.adsabs.harvard.edu/#abs/2019MNRAS.482.5138B} {482, 5138}

\bibitem[\protect\citeauthoryear{{Chattopadhyay} \&
  {Chattopadhyay}}{{Chattopadhyay} \&
  {Chattopadhyay}}{2007}]{2007A&A...472..131C}
{Chattopadhyay} T.,  {Chattopadhyay} A.~K.,  2007, \mn@doi [\aap]
  {10.1051/0004-6361:20066945}, \href
  {https://ui.adsabs.harvard.edu/#abs/2007A&A...472..131C} {472, 131}

\bibitem[\protect\citeauthoryear{{Covino} \& {Pasinetti Fracassini}}{{Covino}
  \& {Pasinetti Fracassini}}{1993}]{1993A&A...270...83C}
{Covino} S.,  {Pasinetti Fracassini} L.~E.,  1993, \aap, \href
  {https://ui.adsabs.harvard.edu/#abs/1993A&A...270...83C} {270, 83}

\bibitem[\protect\citeauthoryear{{Dotter} et~al.,}{{Dotter}
  et~al.}{2010}]{2010ApJ...708..698D}
{Dotter} A.,  et~al., 2010, \mn@doi [\apj] {10.1088/0004-637X/708/1/698}, \href
  {https://ui.adsabs.harvard.edu/\#abs/2010ApJ...708..698D} {708, 698}

\bibitem[\protect\citeauthoryear{{Eggen}, {Lynden-Bell}  \& {Sandage}}{{Eggen}
  et~al.}{1962}]{1962ApJ...136..748E}
{Eggen} O.~J.,  {Lynden-Bell} D.,   {Sandage} A.~R.,  1962, \mn@doi [\apj]
  {10.1086/147433}, \href
  {https://ui.adsabs.harvard.edu/\#abs/1962ApJ...136..748E} {136, 748}

\bibitem[\protect\citeauthoryear{{Eigenson} \& {Yatsyk}}{{Eigenson} \&
  {Yatsyk}}{1989}]{1989SvA....33..280E}
{Eigenson} A.~M.,  {Yatsyk} O.~S.,  1989, \sovast, \href
  {https://ui.adsabs.harvard.edu/#abs/1989SvA....33..280E} {33, 280}

\bibitem[\protect\citeauthoryear{{Ferraro} et~al.,}{{Ferraro}
  et~al.}{2012}]{2012Natur.492..393F}
{Ferraro} F.~R.,  et~al., 2012, \mn@doi [\nat] {10.1038/nature11686}, \href
  {http://adsabs.harvard.edu/abs/2012Natur.492..393F} {492, 393}

\bibitem[\protect\citeauthoryear{{Fraix-Burnet}, {Davoust}  \&
  {Charbonnel}}{{Fraix-Burnet} et~al.}{2009}]{2009MNRAS.398.1706F}
{Fraix-Burnet} D.,  {Davoust} E.,   {Charbonnel} C.,  2009, \mn@doi [\mnras]
  {10.1111/j.1365-2966.2009.15235.x}, \href
  {https://ui.adsabs.harvard.edu/#abs/2009MNRAS.398.1706F} {398, 1706}

\bibitem[\protect\citeauthoryear{{Harris}}{{Harris}}{1996}]{1996AJ....112.1487H}
{Harris} W.~E.,  1996, \mn@doi [\aj] {10.1086/118116}, \href
  {http://adsabs.harvard.edu/abs/1996AJ....112.1487H} {112, 1487}

\bibitem[\protect\citeauthoryear{Kaufman \& Rousseeuw}{Kaufman \&
  Rousseeuw}{1987}]{kaufman1987clustering}
Kaufman L.,  Rousseeuw P.,  1987, Statistical Data Analysis Based on the
  L1-Norm and Related Methods, pp North--Holland

\bibitem[\protect\citeauthoryear{Kaufman \& Rousseeuw}{Kaufman \&
  Rousseeuw}{2008}]{KR}
Kaufman L.,  Rousseeuw P.~J.,  2008, Finding Groups in Data.
John Wiley \& Sons, Inc., \mn@doi{10.1002/9780470316801.fmatter}, \url
  {http://dx.doi.org/10.1002/9780470316801.fmatter}

\bibitem[\protect\citeauthoryear{{Lee}, {Demarque}  \& {Zinn}}{{Lee}
  et~al.}{1988}]{1988csa..proc..149L}
{Lee} Y.-W.,  {Demarque} P.,   {Zinn} R.,  1988, in {Philip} A.~G.~D.,  ed.,
  Calibration of Stellar ages. pp 149--162

\bibitem[\protect\citeauthoryear{{Lee}, {Demarque}  \& {Zinn}}{{Lee}
  et~al.}{1994}]{1994ApJ...423..248L}
{Lee} Y.-W.,  {Demarque} P.,   {Zinn} R.,  1994, \mn@doi [\apj]
  {10.1086/173803}, \href
  {https://ui.adsabs.harvard.edu/#abs/1994ApJ...423..248L} {423, 248}

\bibitem[\protect\citeauthoryear{Lloyd}{Lloyd}{1982}]{lloyd1982least}
Lloyd S.,  1982, IEEE transactions on information theory, 28, 129

\bibitem[\protect\citeauthoryear{MacQueen et~al.}{MacQueen
  et~al.}{1967}]{macqueen1967some}
MacQueen J.,  et~al., 1967, in Proceedings of the fifth Berkeley symposium on
  mathematical statistics and probability. pp 281--297

\bibitem[\protect\citeauthoryear{Maechler, Rousseeuw, Struyf, Hubert  \&
  Hornik}{Maechler et~al.}{2017}]{clusterlibrary}
Maechler M.,  Rousseeuw P.,  Struyf A.,  Hubert M.,   Hornik K.,  2017,
  cluster: Cluster Analysis Basics and Extensions

\bibitem[\protect\citeauthoryear{{Milone} et~al.,}{{Milone}
  et~al.}{2017}]{2017MNRAS.464.3636M}
{Milone} A.~P.,  et~al., 2017, \mn@doi [\mnras] {10.1093/mnras/stw2531}, \href
  {https://ui.adsabs.harvard.edu/\#abs/2017MNRAS.464.3636M} {464, 3636}

\bibitem[\protect\citeauthoryear{{Miocchi}, {Pasquato}, {Lanzoni}, {Ferraro},
  {Dalessandro}, {Vesperini}, {Alessandrini}  \& {Lee}}{{Miocchi}
  et~al.}{2015}]{2015ApJ...799...44M}
{Miocchi} P.,  {Pasquato} M.,  {Lanzoni} B.,  {Ferraro} F.~R.,  {Dalessandro}
  E.,  {Vesperini} E.,  {Alessandrini} E.,   {Lee} Y.-W.,  2015, \mn@doi [\apj]
  {10.1088/0004-637X/799/1/44}, \href
  {http://adsabs.harvard.edu/abs/2015ApJ...799...44M} {799, 44}

\bibitem[\protect\citeauthoryear{{Pasquato} \& {Bertin}}{{Pasquato} \&
  {Bertin}}{2008}]{2008A&A...489.1079P}
{Pasquato} M.,  {Bertin} G.,  2008, \mn@doi [\aap]
  {10.1051/0004-6361:200809462}, \href
  {http://adsabs.harvard.edu/abs/2008A%26A...489.1079P} {489, 1079}

\bibitem[\protect\citeauthoryear{{Pasquato}, {Raimondo}, {Brocato}, {Chung},
  {Moraghan}  \& {Lee}}{{Pasquato} et~al.}{2013}]{2013A&A...554A.129P}
{Pasquato} M.,  {Raimondo} G.,  {Brocato} E.,  {Chung} C.,  {Moraghan} A.,
  {Lee} Y.~W.,  2013, \mn@doi [\aap] {10.1051/0004-6361/201321361}, \href
  {https://ui.adsabs.harvard.edu/#abs/2013A&A...554A.129P} {554, A129}

\bibitem[\protect\citeauthoryear{{Pasquato}, {Miocchi}  \& {Yoon}}{{Pasquato}
  et~al.}{2018}]{2018ApJ...867..163P}
{Pasquato} M.,  {Miocchi} P.,   {Yoon} S.-J.,  2018, \mn@doi [\apj]
  {10.3847/1538-4357/aae52c}, \href
  {https://ui.adsabs.harvard.edu/#abs/2018ApJ...867..163P} {867, 163}

\bibitem[\protect\citeauthoryear{{Recio-Blanco}}{{Recio-Blanco}}{2018}]{2018arXiv181104798R}
{Recio-Blanco} A.,  2018, arXiv e-prints, \href
  {https://ui.adsabs.harvard.edu/#abs/2018arXiv181104798R} {p.
  arXiv:1811.04798}

\bibitem[\protect\citeauthoryear{Rousseeuw}{Rousseeuw}{1987}]{ROUSSEEUW198753}
Rousseeuw P.~J.,  1987, \mn@doi [Journal of Computational and Applied
  Mathematics] {https://doi.org/10.1016/0377-0427(87)90125-7}, 20, 53

\bibitem[\protect\citeauthoryear{{Sandage} \& {Wildey}}{{Sandage} \&
  {Wildey}}{1967}]{1967ApJ...150..469S}
{Sandage} A.,  {Wildey} R.,  1967, \mn@doi [\apj] {10.1086/149350}, \href
  {https://ui.adsabs.harvard.edu/\#abs/1967ApJ...150..469S} {150, 469}

\bibitem[\protect\citeauthoryear{{Searle} \& {Zinn}}{{Searle} \&
  {Zinn}}{1978}]{1978ApJ...225..357S}
{Searle} L.,  {Zinn} R.,  1978, \mn@doi [\apj] {10.1086/156499}, \href
  {https://ui.adsabs.harvard.edu/\#abs/1978ApJ...225..357S} {225, 357}

\bibitem[\protect\citeauthoryear{{Sollima} \& {Baumgardt}}{{Sollima} \&
  {Baumgardt}}{2017}]{2017MNRAS.471.3668S}
{Sollima} A.,  {Baumgardt} H.,  2017, \mn@doi [\mnras] {10.1093/mnras/stx1856},
  \href {https://ui.adsabs.harvard.edu/#abs/2017MNRAS.471.3668S} {471, 3668}

\bibitem[\protect\citeauthoryear{Steinhaus}{Steinhaus}{1956}]{steinhaus1956division}
Steinhaus H.,  1956, Bull. Acad. Polon. Sci, 1, 801

\bibitem[\protect\citeauthoryear{{Zinn}}{{Zinn}}{1985}]{1985ApJ...293..424Z}
{Zinn} R.,  1985, \mn@doi [\apj] {10.1086/163249}, \href
  {https://ui.adsabs.harvard.edu/#abs/1985ApJ...293..424Z} {293, 424}

\bibitem[\protect\citeauthoryear{{Zinn}}{{Zinn}}{1993}]{1993ASPC...48...38Z}
{Zinn} R.,  1993, in {Smith} G.~H.,  {Brodie} J.~P.,  eds, ~ Vol. 48, The
  Globular Cluster-Galaxy Connection. p.~38

\bibitem[\protect\citeauthoryear{{van den Bergh}}{{van den
  Bergh}}{1967}]{1967AJ.....72...70V}
{van den Bergh} S.,  1967, \mn@doi [\aj] {10.1086/110203}, \href
  {https://ui.adsabs.harvard.edu/\#abs/1967AJ.....72...70V} {72, 70}

\makeatother
\end{thebibliography}

\bsp
\label{lastpage}
\end{document}